\documentclass[a4paper,11pt]{article}
\pdfoutput=1 
\usepackage{jinstpub}
\usepackage{subfigure}

\usepackage{textcomp,gensymb} 
\usepackage{multirow}
\hypersetup{breaklinks,colorlinks,allcolors=blue}

\title{\boldmath Recoil Proton Telescopes and Parallel Plate Avalanche Counters for the $^{235}$U(n,f) cross section measurement relative to H(n,n)H between 10 and 450 MeV neutron energy}

\author[a,b,1]{A.~Manna,\note{Corresponding author.}}
\author[c]{E. Pirovano}
\author[d]  {O.~Aberle} 
\author[e]  {S.~Amaducci}
\author[e,d]  {M.~Barbagallo}
\author[g]  {D.M. Castelluccio}
\author[f]  {N.~Colonna}
\author[g]  {P.~Console Camprini}
\author[e]  {L.~Cosentino}
\author[c]  {M.~Dietz}
\author[c]  {Q.~Ducasse}
\author[e]  {P.~Finocchiaro}
\author[h]  {C.~Le Naour}
\author[g,b]  {S.~Lo Meo}
\author[f,i]  {M.~Mastromarco}
\author[a,b]{C.~Massimi}
\author[f,b]  {A.~Mengoni}
\author[l]  {P.M.~Milazzo}
\author[d]  {F.~Mingrone}
\author[c]  {R.~Nolte}
\author[e]  {M.~Piscopo}
\author[c]  {D.~Radeck}
\author[a,b]  {M.~Spelta}
\author[h,d]  {L.~Tassan-Got}
\author[m]  {N.~Terranova}
\author[b]  {G.~Vannini}

\affiliation[a]{Dipartimento di Fisica e Astronomia, Universit\`{a} di Bologna, Italy}
\affiliation[b]{Istituto Nazionale di Fisica Nucleare (INFN), Sezione di Bologna, Italy}
\affiliation[c]{Physikalisch-Technische Bundesanstalt (PTB), Braunschweig, Germany}
\affiliation[d]{European Organization for Nuclear Research (CERN), Switzerland}
\affiliation[e]{INFN Laboratori Nazionali del Sud, Catania, Italy}
\affiliation[f]{Istituto Nazionale di Fisica Nucleare (INFN), Sezione di Bari, Italy}
\affiliation[g]{Agenzia nazionale per le nuove tecnologie (ENEA), Bologna, Italy}
\affiliation[h]{Universit\'{e} Paris-Saclay, CNRS/IN2P3, IJCLab, Orsay, France}
\affiliation[i]{Dipartimento Interateneo di Fisica, Universit\`{a} degli Studi di Bari, Italy}
\affiliation[l]{Istituto Nazionale di Fisica Nucleare (INFN), Sezione di Trieste, Italy}
\affiliation[m]{Agenzia nazionale per le nuove tecnologie (ENEA), Frascati, Italy}

\collaboration[c]{on behalf of the n\_TOF collaboration}

\emailAdd{almanna@bo.infn.it}

\abstract{With the aim of measuring the $^{235}$U(n,f) cross section at the n\_TOF facility at CERN over a wide neutron energy range, a detection system consisting of two fission detectors and three detectors for neutron flux determination was realized. The neutron flux detectors are Recoil Proton Telescopes (RPT), based on scintillators and solid state detectors, conceived to detect recoil protons from the neutron-proton elastic scattering reaction.
This system, along with a fission chamber and an array of parallel plate avalanche counters for fission event detection, was installed for the measurement at the n\_TOF facility in 2018, at CERN. 

An overview of the performances of two RPTs - especially developed for this measurement - and of the parallel plate avalanche counters are described in this article. In particular, the characterization in terms of detection efficiency by Monte Carlo simulations and response to neutron beam, the study of the background, dead time correction and characterization of the samples, are reported. The results of the present investigation show that the performances of these detectors are suitable for accurate measurements of fission reaction cross sections in the range from 10 to 450~MeV.}


\begin{document}
\maketitle
\flushbottom

\section{Introduction}
\label{sec:Intro}
The cross section for $^{235}$U(n,f) is adopted as a standard at thermal neutron energy and at higher neutron energies, between 0.15~MeV and 200~MeV. In addition, the cross section integral in the energy range between 7.8 and 11~eV, is also considered a standard~\cite{Carlson}. 
Therefore, the neutron induced fission of $^{235}$U is extensively used as a reference for measurements of neutron-induced reaction cross sections, for fundamental nuclear physics studies as well as for many applications. For example, in the energy range between 20~MeV and 200~MeV, the $^{235}$U(n,f) reaction is the main reference for neutron fluence measurements in the investigation of biological effects of high-energy neutrons~\cite{Newhauser2011}. High-energy neutron-induced reaction processes are of significant relevance for design studies and development of accelerator-driven nuclear systems~\cite{ADS_FR}. 

Despite its widespread use, however, the recommended $^{235}$U(n,f) cross section data at energies above 20~MeV are based on a very small set of measurements. Hence, there is a clear and long-standing request by the International Atomic Energy Agency (IAEA) to provide new experimental data of $^{235}$U(n,f), relative to n-p scattering, which is considered the primary standard for neutron measurements~\cite{Carlson}. New measurements of absolute cross sections of $^{235}$U reactions in the energy range between 100 to 450~MeV are also included in the NEA high priority request list~\cite{2020EPJWC.23915005D}.
At present no experimental data on neutron induced fission of $^{235}$U above 200~MeV can be found in literature and at these energies, one can only rely on theoretical estimates~\cite{Meo}. 

To deal with this situation, in 2018 an experiment was performed at the n\_TOF facility at CERN, to measure the fission cross section of $^{235}$U relative to n-p elastic scattering, featuring a redundant apparatus consisting of two chambers to measure fission events and three detectors to determine the incident neutron flux.
The combination of different detectors and techniques allowed to cross check the results and better estimate the systematic uncertainty of this challenging measurement. 

In this paper, the two multi-stage recoil proton telescopes developed for this experiment and the parallel plate avalanche counters (PPACs) detectors are also referred as the ‘high-energy’ setup and the 3-stage recoil proton telescope and parallel plate fission-ionization chamber (PPFC) compose the ‘low-energy’ setup.
The focus of this article is on a thorough description and characterization of the high energy setup, while a description of the low energy setup is provided in ref.~\cite{Pirovano}. 
This article consists of three main sections. In section~\ref{sec:measurement} an overview of the n\_TOF facility and of the experimental setup used in the measurement campaign is provided. Section~\ref{sec:ms_rpt} describes the two multi-stage recoil proton telescopes followed by the study of the efficiency and the identification of the background components involved in the data analysis, performed using Monte Carlo simulations. In section~\ref{sec:ppac}, the three parallel plate avalanche counters and the $^{235}$U samples used are characterised, with a focus on the efficiency calculation.

\section{The measurement}
\label{sec:measurement}
\subsection{The n\_TOF facility} 
While several quasi-monoenergetic neutron sources with energies up to 400~MeV are available worldwide~\cite{IAEA2014}, measurements of neutron induced reaction cross sections covering a continuous wide neutron energy spectrum are rare.
n\_TOF at CERN is one of the few facilities offering the opportunity to perform cross section measurements in a broad neutron energy range, all the way from thermal energies up to 1~GeV.
The n\_TOF pulsed neutron source is based on a 20~GeV/c proton beam from the CERN Proton Synchrotron accelerator (PS) impinging on a cylindrical lead target of 60 cm diameter and 40 cm length.  
Protons are accelerated in high-intensity bunches of 7~ns (rms) time width and repetition rate of less than 1~Hz~\cite{Gunsing2007}.
The proton beam can be delivered on target in two different operational modes, dedicated and parasitic, differing from each other mainly in terms of proton intensity: about 7×10$^{12}$ and 3.5×10$^{12}$ protons per bunch, respectively.
The experiment described here was performed in the first experimental area EAR1~\cite{Guerrero}, at the nominal distance of 185~m from the spallation target using the time-of-flight technique. Overall, the energy resolution was excellent (10$^{-4}$ $<$ $\Delta$E/E $<$10$^{-2}$), thanks to the long neutron flight path leading to EAR1.     

\subsection{Experimental setup}
Because of the challenges present in these type of experiments and the wide energy region to cover, several detectors were used at the same time. With the aim of benchmarking our detection setup and control systematic uncertainties, we adopted a redundant approach.
To start with, fission events were detected by two chambers.
(59.8\,$\pm$\,0.3)~mg of $^{235}$U was placed in the beam and stacked in 10 samples: (27.11\,$\pm$\,0.16)~mg contained in the PPACs and (32.660\,$\pm$\,0.001)~mg contained in the PPFC.
Thanks to the fast time response and the very low sensitivity to the $\gamma$-rays, the PPAC detector has already been successfully used at n\_TOF to measure the fission cross section (relative to $^{235}$U) of a large number of actinides up to 1~GeV neutron energy~\cite{Paradela_2010, Tarrio_2010}. However, the PPAC detects only fission fragments emitted into a forward cone with an opening angle of about 60$^\circ$ and its fragment detection efficiency is not straightforward to evaluate (detailed discussion in the section~\ref{sec:PPAC_eff}). On the other hand, the detection  efficiency of the PPFC detector is very well characterized. Its drawback is the limited energy region where the chamber can operate. In fact, due to the background induced by $\gamma$-rays from the spallation target, it is limited to energies below approximatively 200~MeV. To complement these drawbacks, in the present experiment it has been decided to utilize both detection systems at the same time.

Prerequisite for this measurement was the availability of a suitable detector able to measure the neutron flux at energies above 20~MeV.
For this purpose, we developed three Recoil Proton Telescopes (RPTs) consisting of a radiator (a polyethylene target) and a multi-detector positioned at a small angle ($\theta$) with respect to the neutron beam direction to detect the recoil protons from n-p scattering events.
Examples of RPTs can be found in the literature for neutron energies below 70~MeV~\cite{Donzella, Caiffi}, and up to 200~MeV~\cite{Dangendorf}. 
A 3-stage recoil proton telescope (3S-RPT), composed by two transmission detectors and one stop detector was used, exploiting the approach of ref.~\cite{Dangendorf}.

In order to extend the energy range a new compact design of a multi-stage recoil proton telescope (MS-RPT) was implemented, based on similar concepts but different technical details, capable of reaching up to 450~MeV neutron energy. 
As will be shown below, PPACs in combination to MS-RPTs are demonstrated to be detectors well suited to measure the $^{235}$U fission cross section from 10 MeV up to 450~MeV. The upper limit is defined by the neutron flux measurement. In fact, although it is possible to recognise and extract the properties of events generated by neutrons up to 1~GeV, the total neutron-proton scattering cross section is completely elastic only up to about 300 MeV, where the first inelastic channel
\begin{center}
$n + p \longrightarrow n + \Delta^+ \longrightarrow n + p + \pi^0$ \qquad Q$_{value}$ = 293~MeV/c$^2$\\
\end{center}
opens. This reaction process becomes relevant above 450~MeV and remains significant up to 1~GeV, when it reaches about 40\%. In particular, the inelastic cross section at 350~MeV neutron energy is 0.5\% of the total n-p scattering cross section and at 450~MeV it reaches a few percent~\cite{Bystricky}. 
Therefore, as the reference cross section used includes only the n-p elastic scattering channel, a correction as been applied to our data by subtracting the contribution of the inelastic channel which amount to a maximum of 3.8\% at 450~MeV remaining always below the elastic channel cross-section uncertainty declared by Arndt~\cite{INDC1997, 1992PhRvD..46.1192A}.
Consequently, the neutron flux extraction is achievable in the neutron energy range between 10 and 450~MeV. \\


\begin{figure}[ht]
    \centering
    \includegraphics[width=0.75\textwidth]{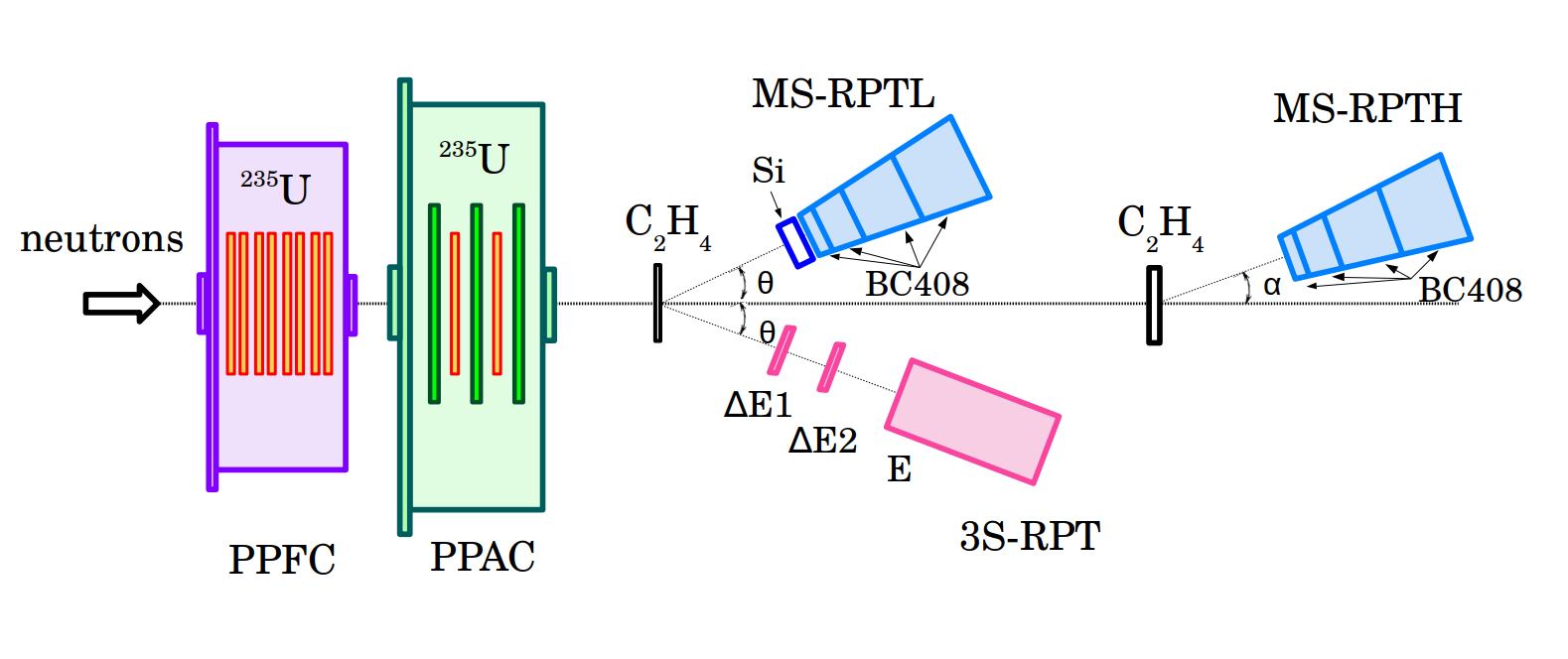}
\caption{Setup for the measurement of the $^{235}$U(n,f) cross section. Fission events were measured with a Parallel Plate Ionization Chamber and an array of Parallel Plate Avalanche Counters. Recoil protons were detected and identified by three different telescopes, placed at small angles out of the neutron beam.}
    \label{fig:SetUp} 
\end{figure}

Figure~\ref{fig:SetUp} shows the scheme of the final experimental setup, including the two chambers to count the fission events from the uranium targets and three recoil proton telescopes to measure, simultaneously, the neutron flux. 
Downstream of the fission chambers, the two hydrogen-compound samples, mounted along the neutron beam, act as targets for the proton-neutron elastic scattering reaction.
The 3S-RPT and MS-RPTL were positioned symmetrically at $\pm$25$^\circ$ with respect to the neutron beam direction. 
Such an angle was chosen as a trade-off between the need of a reasonable counting statistics and the need to place the detectors outside the primary neutron beam. 
In order to maximize the hydrogen density in the solid target, polyethylene (C$_2$H$_4$) samples were chosen. Three different sample thicknesses were employed, namely 1~mm, 2~mm and 5~mm, in dedicated runs optimized for different neutron energy ranges. MS-RPTH was installed 50~cm downstream, facing a 5~mm thick polyethylene sample and positioned at 20$^\circ$ with respect to the beam direction. This dedicated telescope was conceived for neutron energies above 120~MeV. \\
The measurement of the neutron fluence consisted in detecting, identifying and counting the protons elastically scattered by the polyethylene samples. 
Although this material presents a favorable stoichiometric ratio between hydrogen and carbon, the latter is a source of background that has to be estimated (and subtracted) by means of dedicated measurements. To this purpose, data were acquired with graphite samples of corresponding areal density of carbon atoms, in order to subtract the contribution of the n+C reactions from the polyethylene data. 
The different samples, and their main features, used during the experimental campaign are presented in table~\ref{tab:sample_PRT}.
A combustion analysis was performed at the ZEA-3 unit at the Forschungszentrum Jülich (Germany) to determine the stoichiometric proportion of the number of hydrogen and carbon atoms~\cite{Pirovano}.

 \renewcommand\arraystretch{1.3}
 \begin{table*}[h]
 \centering
 \scriptsize
 \caption{Characteristics of the samples.}
 \centering
\begin{tabular}{cccccc} 
& & & & \\
 Sample & Name & Thickness  & Mass & Areal density \\
 & & mm & g & g/cm$^2$ \\
\hline
\hline
& 1 mm & 1.025\,$\pm$\,0.004 & 9.761\,$\pm$\,0.005 & 0.0978\,$\pm$\,0.0004  \\
C$_2$H$_4$ & 2 mm & 1.824\,$\pm$\,0.011 & 17.240\,$\pm$\,0.005 & 0.1743\,$\pm$\,0.0011  \\
& 5 mm & 4.925\,$\pm$\,0.004 & 47.193\,$\pm$\,0.005 & 0.4726\,$\pm$\,0.0011  \\
\\[-1.5em]
\hline
\\[-1.35em]
& 0.5 mm & 0.500\,$\pm$\,0.004 & 9.066\,$\pm$\,0.005 & 0.0887\,$\pm$\,0.0008  \\
C & 1 mm & 1.000\,$\pm$\,0.005 & 17.480\,$\pm$\,0.005 & 0.1736\,$\pm$\,0.0012 \\
& 2.5 mm & 2.500\,$\pm$\,0.004 & 44.103\,$\pm$\,0.005 & 0.4378\,$\pm$\,0.0011\\
\hline
\end{tabular}
\label{tab:sample_PRT}
\end{table*}
\renewcommand\arraystretch{1.2}

Considering the proton energy range under of interest the energy loss in air is negligible.  

\subsubsection{Alignment of the setup in the beam}
The alignment of the detection system with respect to the neutron beam is crucial for the RPTs. Indeed, on one hand $\vartheta_p$ (the angle subtended by the detector) is necessary for the differential elastic scattering cross section to be considered, d$\sigma$/d$\Omega_p$, while it is needed also to accurately evaluate the efficiency of the RPTs.
Therefore, the exact angle and the positions of the three telescopes were determined by means of the Leica Absolute Tracker AT401~\cite{Laica}, composed by a laser source and a reflector. 
The absolute distances of each element from the laser source were measured in a 3D coordinate system and their positions were evaluated using the coordinates system with respect to the geometrical neutron beam line. 
The transversal neutron beam position and the direction had been identified, with a resolution of a tenth of a millimeter, by two Quads Timepix detectors~\cite{George}, produced by the Medipix2 collaboration at CERN. Then, the positions of the two polyethylene targets have been carefully calculated.
The target-RPT distances and the angles subtended by the RPTs (indicated as $\theta$ and $\alpha$ in Figure~\ref{fig:SetUp}) are 15.74~cm and 25.07$^\circ$ for MS-RPTL, 21.6~cm and 20.32$^\circ$ for MS-RPTH, respectively. The accuracy of all the measured position is 0.15~mm (one sigma).

\section{The multi-stage telescope approach}
\label{sec:ms_rpt}
\begin{figure}[ht]
    \centering
    \includegraphics[width=0.45\textwidth]{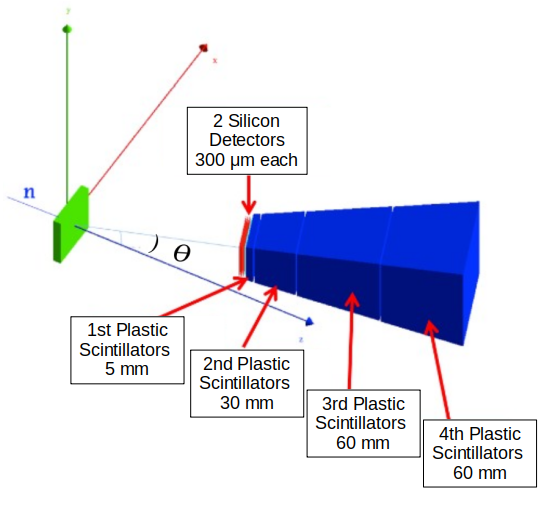}
\caption{Geometrical drawing of the multi-stage telescope: the two silicon detectors followed by the four plastic scintillators.}
    \label{fig:MS-RPT_s}
\end{figure}

The two multi-stage recoil proton telescopes (MS-RPTL and MS-RPTH), developed by the Istituto Nazionale di Fisica Nucleare (INFN), were designed to be as compact as possible with a slender and light mechanical structure to minimize the background. A 3D sketch of the full version and its basic geometrical drawing are shown in Figure~\ref{fig:MS-RPT_s}.

Both MS-RPTs consist of a trapezoidal structure pointing to the polyethylene sample, made of four slabs of BC408 plastic scintillator~\cite{saint-gobain} (hereafter referred to as sub-detectors: P1,P2, P3 and P4) with increasing thickness of 0.5, 3, 6, and 6~cm, respectively. The total length of the two detectors is 16~cm with an increasing transverse size, from 3$\times$3~cm$^2$ to 7.2$\times$7.2~cm$^2$, thus covering a constant solid angle along the axis. 
Each sub-detector is read-out independently, with each one or a combination of two or more of them, acting either as $\Delta$E or E detector. In this way, it has been possible to discriminate recoil protons from other light charged particles, e.g. deuterons and tritons, emitted in the interaction of neutrons with the polyethylene sample, and possibly to determine their energy. 
The thick scintillators P2, P3, P4 are coupled to a Ø = 25~mm Hamamatsu R1924A photo multiplier tube~\cite{pmt_datasheet} at the center of a side face. 
In order to minimize the dependence of the light collection on the impact point, the thinnest scintillator, P1, was coupled to two PMTs on opposite sides by means of light guides. 
The pulse output of P1 has been calculated as the geometrical average between the two recorded signals, a technique already successfully employed in other applications~\cite{Agodi}. The sub-detectors were first wrapped with 50~$\mu$m teflon and then with 4~$\mu$m aluminized mylar, in order to maximize the scintillation light collection efficiency, while preventing cross talks between adjacent detectors. Finally, the scintillators were put together by means of adhesive aluminum tape and installed onto a very light aluminum mechanical support fixed on a PET basement.
The MS-RPTL, dedicated to the low neutron energy range of the measurement, had the same structure but with two silicon detectors (indicated by S1 and S2) placed in front of the plastic scintillators. They were used in order to decrease the minimum energy of detected neutrons in the n-p interaction process, from 40~MeV for the bare scintillators, to 10 MeV, thus allowing to check the deduced cross section in a region where reliable data already exist. The fully depleted silicon detectors S1 and S2 are micron semiconductor 300~$\mu$m thick silicon detectors MSX09-300~\cite{silicon_datasheet}, with an active area of 3$\times$3~cm$^2$. They were installed, at a distance of 7~mm from each other, inside a dedicated case shielded against electromagnetic noise with a 4~$\mu$m aluminum foil on the entrance and exit windows. An additional 20~$\mu$m aluminum cap was also installed as a redundant protection.
The front-end electronics of the two silicon detectors consisted of the charge preamplifier NPA-16 FL (NeT Instruments) and the ORTEC 474 timing filter amplifier module.

A problem of TOF measurements is related to the $\gamma$-flash, i.e. the large prompt signal produced by relativistic particles and $\gamma$-rays from the spallation process in the neutron production site, often blinding all the detectors for some time after their arrival in the experimental area~\cite{Guerrero}. Consequently, in TOF measurements it is preferable to use fast detectors with a fast recovery time, reducing the saturation time after the $\gamma$-flash, thus allowing to extend the measurable neutron energy range to higher values. 

\begin{figure}[ht]
\centering
\includegraphics[width=0.85\textwidth]{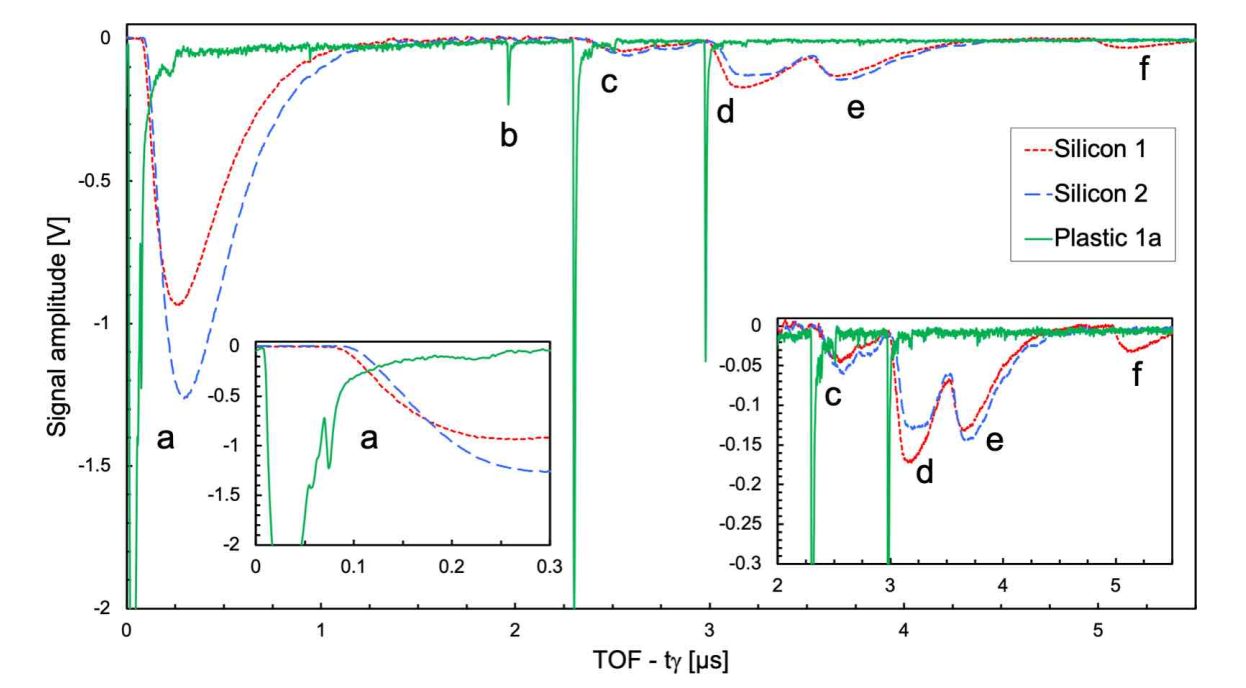}
\caption{Sample signals produced by the two silicon detectors, shown in blue and red. The signal from the first plastic scintillator is shown in green.}
\label{fig:movie}	
\end{figure}

As an example of detector output observed online during the measurement, a snapshot of a digitized data stream from the S1, S2 and P1 detectors of the MS-RPT telescope is shown in Figure~\ref{fig:movie}. In the region (a), we observe the $\gamma$-flash signal, which is rather long for the silicon detectors ($\sim$\,1$\mu$s) and quite fast for the P1 plastic scintillator  ($\sim$~\,100~ns). In particular, on the tail of the P1 $\gamma$-flash, a useful signal from a neutron of about 1.3~GeV is present, which can be resolved and analyzed by means of the dedicated n\_TOF Pulse Shape Analysis code~\cite{Zugec}. \\

The analysis of the events in coincidence between two or more adjacent active scintillator slabs is the principle at the base of the events selection. In fact, the coincidence contributes to the suppression of the $\gamma$ background and guarantees the identification of the events (and relative particles) originating only from the polyethylene sample. Moreover, thanks to the multi-stage structure it is possible to select the best configuration of sub-detectors in coincidence in the different neutron energy range. In figure~\ref{fig:movie}, there are a few examples of detector outputs for different neutron energies. In region (b), a proton produces a signal in P1 but leaves too little energy in S1 and S2 to be detected. In (c) a valid signal is detected in S1, S2 and P1, while the region (d) shows a still slower proton leaving more energy in S1 and S2 and less in P1. Region (e) represents a case where the proton is so slow that stops in S2. Finally, region (f) shows the case of a very low energy proton that is stopped in S1. 
We can conclude that the multistage structure allowed to work with a specific combination of sub-detectors in coincidence, depending on the energy range being analysed. Each combination of scintillators in coincidence is effective until the proton stops inside the last layer of the detector taken into account. Protons energy higher than the so-called punch-through energy, reach the next layer of the RPT which, therefore, has to be included in the coincidence for the events selection.

In Table~\ref{tab:E_loss} we list the expected energy deposited by protons originated in the sample and entering the telescope perpendicularly to its entrance face. The proton energy values, and the corresponding neutron energy values, were chosen very close to the punch-through energy for each telescope layer. 

\begin{table*}[h]
    \centering
        \caption{Energy deposited into the six MS-PRTL layers by protons impinging perpendicularly with
several energies close to the punch-through across each detector element.}
    \begin{tabular}{c c c c c c c}
    E$_{neutron}$&  E$_{dep}$ S1&  E$_{dep}$ S2&  E$_{dep}$ P1&  E$_{dep}$ P2&  E$_{dep}$ P3&  E$_{dep}$ P4\\
    MeV & MeV & MeV & MeV & MeV & MeV & MeV \\
    \hline
    \hline
         
183 & 0.3 & 0.3 & 2.8 & 17.9 & 43.0 & 84.7 \\
140 & 0.4  & 0.4 & 3.4 & 22.7 & 87.3 & - \\
82 & 0.6 & 0.6 & 5.4 & 59.8 & - & - \\
30 & 1.2 & 1.3 & 21.8 & - & - & - \\
10 & 3.7 & 4.0 & - & - & - & - \\
7 & 5.7 & - & - & - & - & - \\
\hline
    \end{tabular}
    \label{tab:E_loss}
\end{table*}

\begin{figure}[ht]
    \centering
    \subfigure[]{\includegraphics[width=0.48\textwidth]{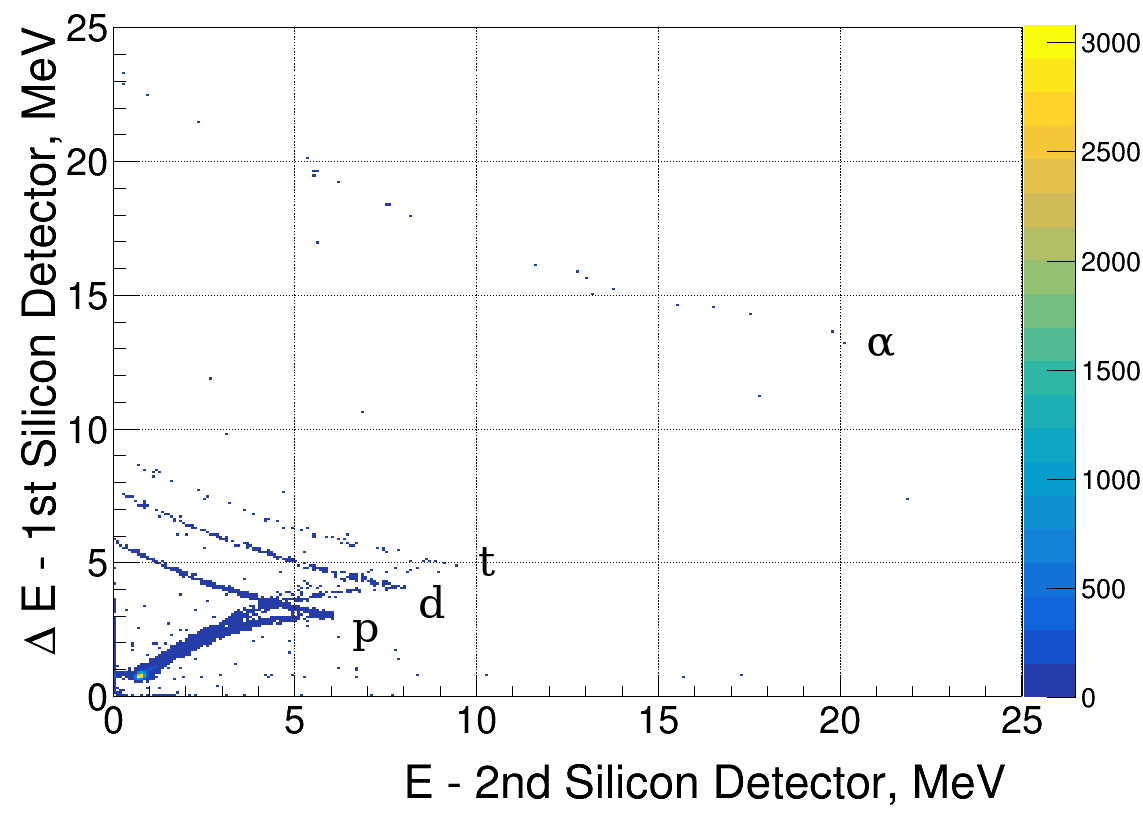}
    \label{DEE_54MeV_sili}}	
        \subfigure[]{\includegraphics[width=0.48\textwidth]{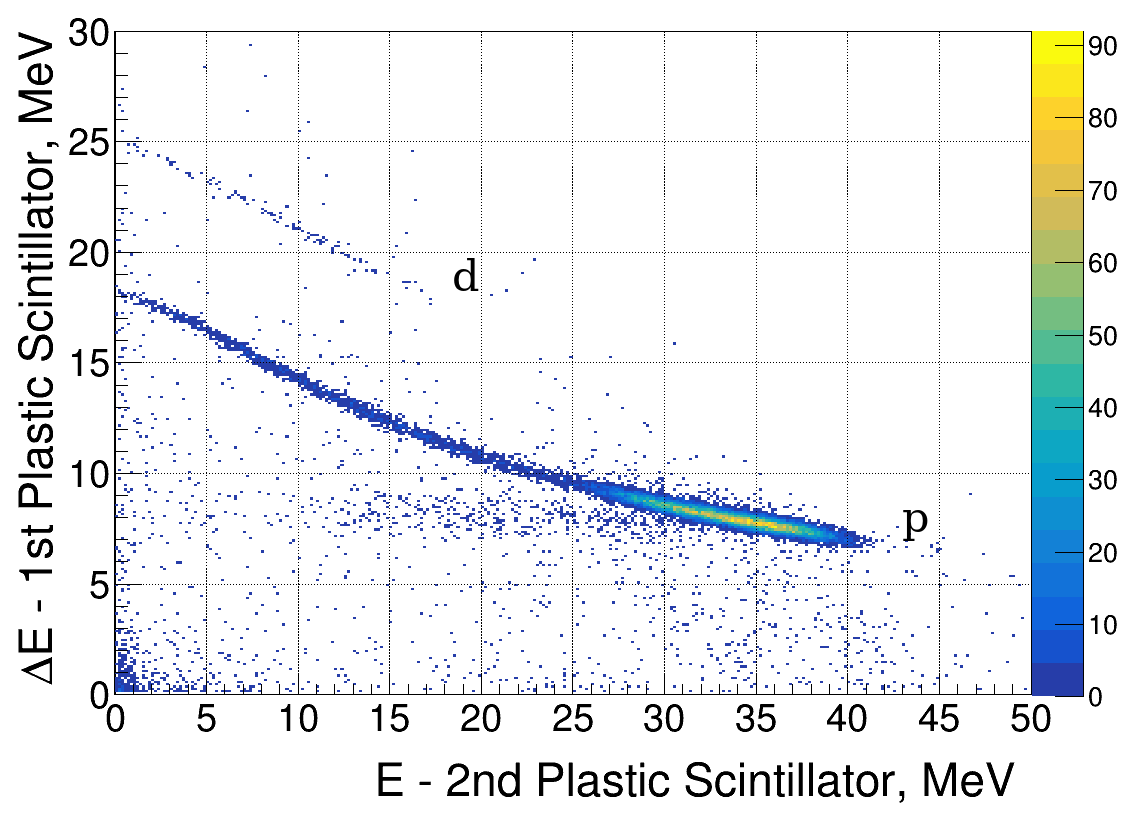}
        \label{DEE_54MeV_Scint}}	
\caption{Monte Carlo simulation of a 56 MeV neutron energy beam impinging on a 2-mm thick polyethylene target. Figure~\ref{DEE_54MeV_sili} shows the energy deposited by protons, deuterons, tritons and $\alpha$ in the first silicon detector, $\Delta$E, as a function of the energy deposited in the second silicon detector, E. The $\Delta$E-E matrix~\ref{DEE_54MeV_Scint} shows the energy deposited by protons, deuterons in the first scintillator as a function of the energy deposited by protons and d in the second one.}
\label{fig:DE_E_seg}
\end{figure}

As described in the next section, detailed Monte Carlo simulations were performed in order to design and study the detectors' feature. For instance, Figure~\ref{fig:DE_E_seg} shows how it is possible to identify the background events coming from $^{12}$C (deuterons, tritons and alpha particles), using two $\Delta$E-E matrices produced by simulated data of a 56~MeV neutron energy beam impinging on a polyethylene target. In Figure~\ref{DEE_54MeV_sili} $\Delta$E and E are the energy deposited in the first and second silicon detector, respectively. In this configuration it is possible to identify four hyperbolas produced by four different families of particles: protons, deuterons, tritons and $\alpha$. Note that only the latter are stopped in the second silicon detector. On the contrary, taking into account the particles detected in the first two plastic scintillators (Figure~\ref{DEE_54MeV_Scint}), only protons and deuterons remain, since the $\alpha$ particles at that energy are stopped in the second silicon detector and the tritons in the first scintillator. Therefore, in this configuration the selection of events generated only by protons is straightforward, just selecting the contours around the hyperbola generated by protons.

\subsection{Monte Carlo simulations}
\label{sec:MC}
Extensive simulations with neutrons impinging on a realistic setup, composed by either the polyethylene or the carbon samples and the recoil proton telescopes, were performed to calculate the detector's response. 
The efficiency of the telescope depends on several elements: the geometrical configuration, the multiple scattering of the particles in the sample and in the telescope itself, the effect of thresholds and analysis conditions applied for events selection. In the simulations, the geometry of the whole experimental setup was implemented in the Monte Carlo (MC) simulations, and the physical processes involved in the particle detection were properly taken into account.
The same analysis approach was adopted in both experimental data and simulations, by constructing the $\Delta$E-E matrices and applying the same analysis criteria to discriminate between background and events produced by n-p elastic scattering. 

Monte Carlo simulations were carried out with two toolkits: Geant4~\cite{Agostinelli} and MCNP~\cite{Briesmeister}. This dual approach was chosen as the efficiency determination of the RPTs is based exclusively on MC simulations.
A step-by-step comparison of the results from the two toolkits was performed to validate Geant4 and MCNP simulations. 
For each step it was verified that the two codes provided the same results after the adoption of the correct physics models for the n-p elastic scattering~\cite{Terranova}. In particular, the energy deposited in each stage of the counter telescope and the number of recorded events resulted to be compatible, within the uncertainties, between the two MC simulations. For instance, a (50\,$\pm$\,1.2)~MeV neutron beam produced a number of events, identified as from n-p scattering in a ratio of 1.013 for the simulation performed with MCNP and GEANT4. An article with detailed description of the simulation results is in preparation~\cite{Ducasse2023}. 
This agreement allowed us to gain confidence on the detection efficiency obtained from the simulations.
The first item considered was the response of the RPT to both point-like and extended sources of monoenergetic protons, deuterons, and $\alpha$ particles. 
Once these conditions were verified, a more detailed and complete simulation was performed. In particular, monoenergetic beams (with the n\_TOF spatial profile) of neutrons were transported through the polyethylene targets, with the same characteristics as those used during the experiment. 
The results obtained with the Geant4 code will be shown, owing to their compatibility with MCNP.

The first component to be studied is the geometrical factor, which defines the solid angle subtended by the detector. Since the area of the two silicon detectors is the same, when the coincidence between the two is imposed, the second one defines the solid angle and the corresponding geometric efficiency is $\varepsilon$ = (0.0298\,$\pm$\,0.0004). For the trapezoid, composed by the four plastic scintillators, coincidences between all sub-detectors are required and the solid angle is determined by the acceptance of the first scintillator, $\varepsilon$\,=\,(0.0308\,$\pm$\,0.0004).
Assuming an isotropic proton source contained in a volume, determined by the beam profile size and placed in the position of the polyethylene target, it is possible to derive the geometric efficiency of the whole telescope, taking into account the possibility of multiple scattering of the protons in the scintillating material. In fact, protons can lose energy in the first plastic scintillator, reach the second one and, following an interaction, escape from the telescope. In this situation, the event locates off the proton hyperbola in the $\Delta$E-E matrix. Although generated by n-p scattering, such event is badly identified and is therefore lost. For this reason it is important to quantify with precision the effect of multiple scattering on the total efficiency of the detector. 

\begin{figure}[ht]
\centering
\subfigure[]{\includegraphics[width=0.45\textwidth]{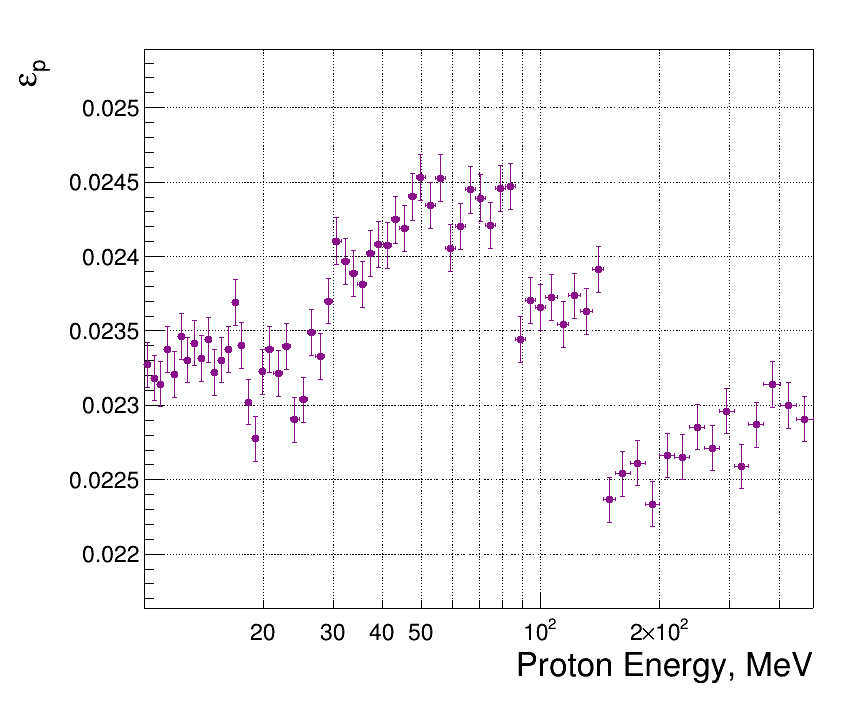}
\label{fig:eff_p}}
\subfigure[]{\includegraphics[width=0.45\textwidth]{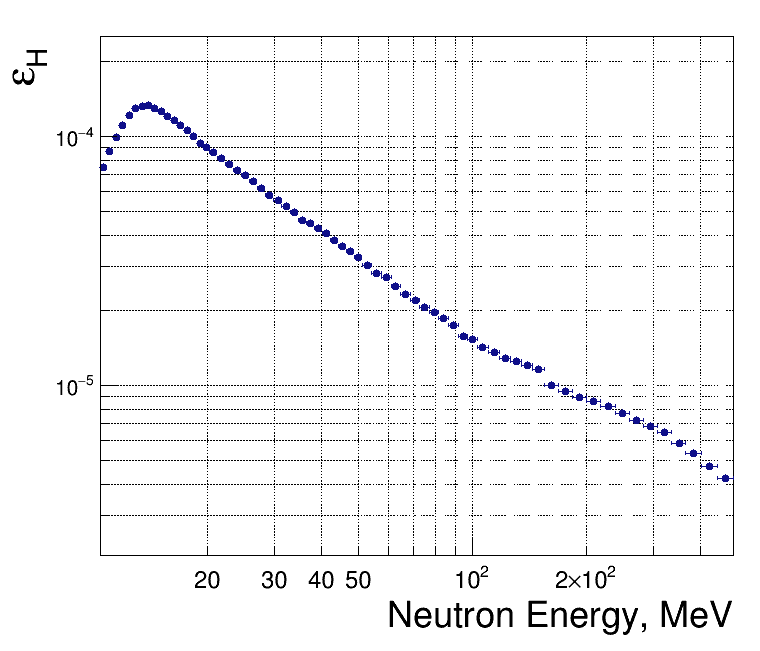}
\label{fig:eff_H}}
\caption{In the left panel the efficiency is studied through an isotropic proton source placed in the polyethylene sample position. The efficiency thus calculated incorporates the evaluation of the solid angle subtended by the telescope and the effect due to multiple scattering suffered by protons in the target and in the detector itself. In the right panel the fraction of protons detected by the telescope divided by total number of neutrons hitting an hydrogen sample with a thickness and an areal density of 0.384~mm and 0.91~g/cm$^2$ is shown.}
\label{fig:effi_p_H}
\end{figure}

Figure~\ref{fig:eff_p} shows the ratio between the number of detected events, divided by total number of generated protons (10$^6$) as a function of the proton energy. The transition among the different configurations of detectors in coincidence, at about 30, 90 and 150~MeV, are reflected in the variations in the efficiency curve.

The second step towards the final simulation of the complete experimental setup, was achieved assuming an hydrogen sample, 0.384~mm thick and an areal density of 0.91~g/cm$^2$. In this simulation, the neutron beam interaction with the target was included. The neutron beam profile has been experimentally determined for the EAR-1 at n\_TOF: a gaussian-like profile characterized by a standard deviation of 0.56~cm~\cite{Belloni,Pancin}. The result of the MC simulation is shown in Figure~\ref{fig:eff_H}. The efficiency is calculated with the neutron source impinging on a hydrogen target. With the addition of the sample, the effects produced by the target itself, which are mainly the energy deposited in it and the multiple scattering, are included in the efficiency study.

The differential elastic scattering n-p cross section is a necessary ingredient needed to perform the simulations.
The VL40 phase-shift energy-dependent solution, obtained by Arndt and collaborators~\cite{Arndt_83, Arndt_87}, is the recommended interaction for the n-p scattering cross section. This evaluation is accepted by the the Committee on Standards NEANDC/INDC as a primary standard for cross section measurements in the 20-350~MeV range~\cite{Carlson}, and included in the evaluated nuclear data files ENDF/B-VII.0 and JENDL-4.0~\cite{CHADWICK20062931, Shibata} evaluations in the low energy range, below 20 MeV, and in the JENDL/High Energy file~\cite{Watanabe} up to 3 GeV.
Since all the reference physics list classes already defined in Geant4 do not contain the VL40 interaction, a detailed Monte Carlo model~\cite{Terranova} was specifically developed for using the Geant4 toolkit~\cite{Agostinelli}.

\subsubsection{Background components}
The simulations were completed by replacing the hydrogen sample with polyethylene. The carbon present in the sample is responsible for the main component of background and manifests itself through various processes.
When neutrons interact with carbon nuclei, different reactions can take place, producing neutrons, protons, deuterons, tritons and $\alpha$ particles. The main reaction channels, with lower thresholds are summarised in table~\ref{tab:C_reaction}. 
\begin{center}
 \renewcommand\arraystretch{1.3}
 \begin{table}[h]
\caption{Main background reactions produced by the interaction of neutrons with carbon.}
\small
\centering
\begin{tabular}{lcc} 
\\[-1.3em]
Reaction & & Q-value (MeV) \\
\\[-1.2em]
\hline
\hline
\\[-1.4em]
n + $^{12}$C $\longrightarrow$ $^9$Be + $\alpha$ & & -\,5.7\\
n + $^{12}$C $\longrightarrow$ n + 3 $\alpha$ & & -\,7.9\\
n + $^{12}$C $\longrightarrow$ $^{12}$B + p & & -\,13.6\\
n + $^{12}$C $\longrightarrow$ $^{11}$B + n + p & & -\,17.3\\
n + $^{12}$C $\longrightarrow$ $^{11}$B + d & & -\,14.9\\
n + $^{12}$C $\longrightarrow$ $^{10}$B + t & & -\,20.5\\
\\[-1.2em]
\hline
\end{tabular}
\label{tab:C_reaction}
\end{table}
 \renewcommand\arraystretch{1.2}
\end{center}
As already described above, the segmentation of the telescope ensures the suppression of the background related to particles with Z$>$1, that are stopped in the first stages of the detector. Therefore, in the $\Delta$E-E matrix, only the hyperbola generated by protons and deuterons remains. Consequently, a condition in the $\Delta$E-E matrix is enough to select the events containing protons.\\

On the other hand, neutron induced reactions on carbon, as well as n-p scattering in hydrogen, can generate neutrons emitted in the direction of the telescope. These neutrons entering in the telescope can cause, in turn, reactions generating protons in the detector itself. This effect becomes more important for the plastic scintillators as they contain hydrogen and carbon. By imposing a time-coincidence from the first scintillator to the last one, this component of background is considerably reduced. Background events produced by neutrons (and other particles) in the second, third and fourth scintillators  are removed. The number of protons produced in the last three plastic scintillators was determined through Monte Carlo simulations. In the case of incident 150-MeV neutrons, about 25\% of the total number of protons recorded in the last stage are due to this mechanism. By requiring the coincidence between scintillators this contribution is reduced to 5\% while, with the proper event selection (example in section~\ref{sec:RPT_response}), this background contribution is reduced to zero. \\
Finally, a not negligible component of background is still present, due to protons directly generated by the neutron interaction with carbon contained in the target. 

\begin{figure}[ht]
\centering
\subfigure[Hydrogen sample]{\includegraphics[width=0.45\textwidth]{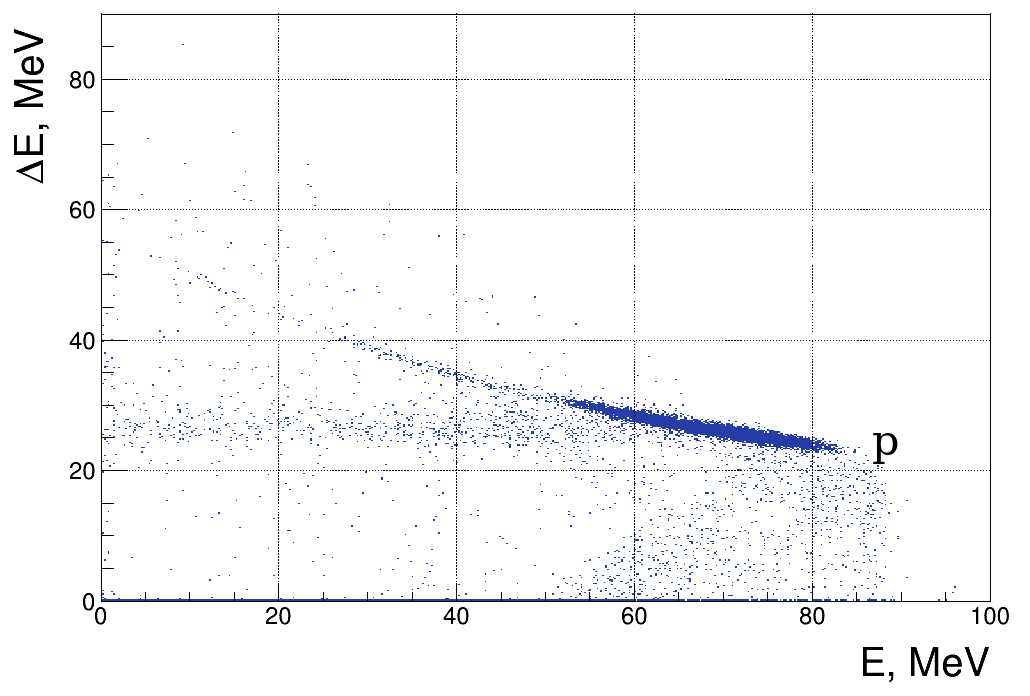}\label{fig:DEE_H_120}}
\subfigure[][Carbon sample]{\includegraphics[width=0.45\textwidth]{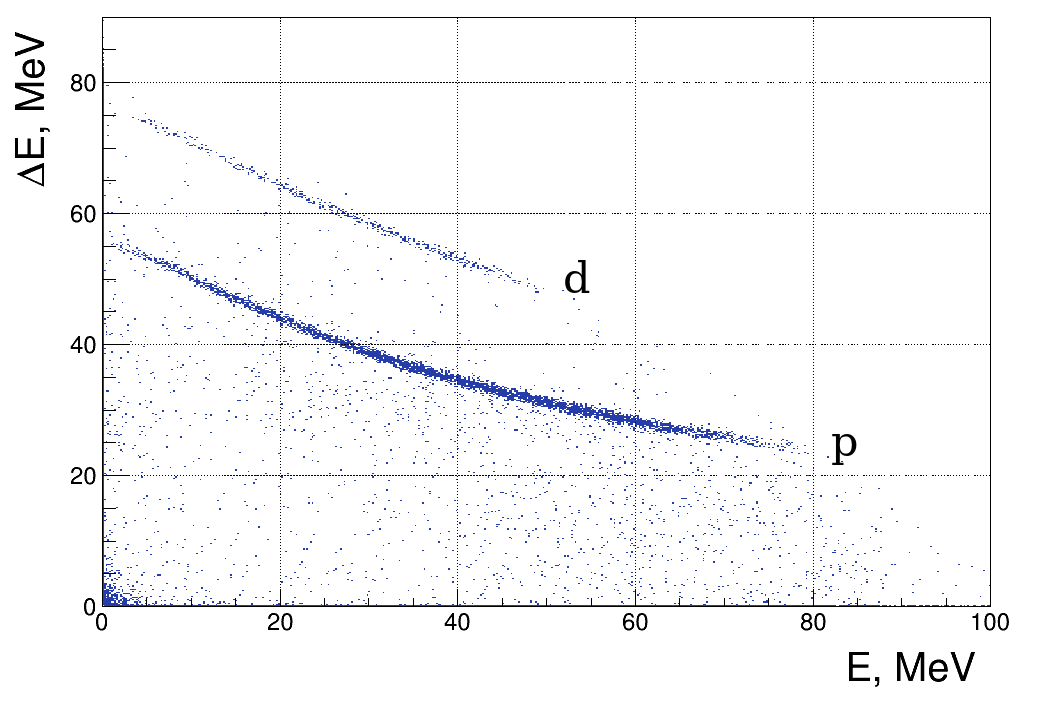}\label{fig:DEE_C_120}}
\caption{$\Delta$E-E matrices are generated by Monte Carlo simulations with a 120-MeV neutron beam.
Figure~\ref{fig:DEE_H_120} shows the output of an H sample placed on the neutron beam. The detected protons can only derive from the n-p scattering reaction, in fact all the events are positioned in a narrow region of the matrix. Figure~\ref{fig:DEE_C_120} displays the $\Delta$E-E matrix from the neutron beam hitting a carbon sample. In this case there are two groups of events related to the families of protons and deuterons; the events are distributed throughout the hyperbola without any peak.}
\label{fig:DE_E_120MeV}
\end{figure}

Figure~\ref{fig:DE_E_120MeV} shows the $\Delta$E-E matrix produced by Monte Carlo simulations, assuming the neutron beam hitting a hydrogen target (on the left panel) and a carbon target (on the right panel). The different types of nuclear reactions involving carbon and hydrogen are clearly separated:\\
\phantom{Ai}i) in the case of n+H, only protons are present in the $\Delta$E-E matrix and their energy is distributed around the corresponding kinematic locus (defined, in the non-relativistic approximation, by E$_p$\,=\,E$_n$\,cos$^2\,\theta$).\\
\phantom{Ai}ii) In the case of n+C, in addition to protons other particles are present. As the Q-value of the neutron-induced reactions on C is negative, the energy distribution of emitted particles has a lower mean value and the kinematics of the nuclear reaction does not produce a peaked distribution in energy.

In summary, all the events produced by the n-p scattering are localized in a restricted region of the matrix, while those produced by the interaction with carbon are distributed throughout the whole hyperbola. \\
For the estimation of the remaining background components, the Monte Carlo simulation was carried out with a carbon sample instead of the polyethylene, as in the actual measurement.

\begin{figure}[ht]
\centering
\includegraphics[width=0.75\textwidth]{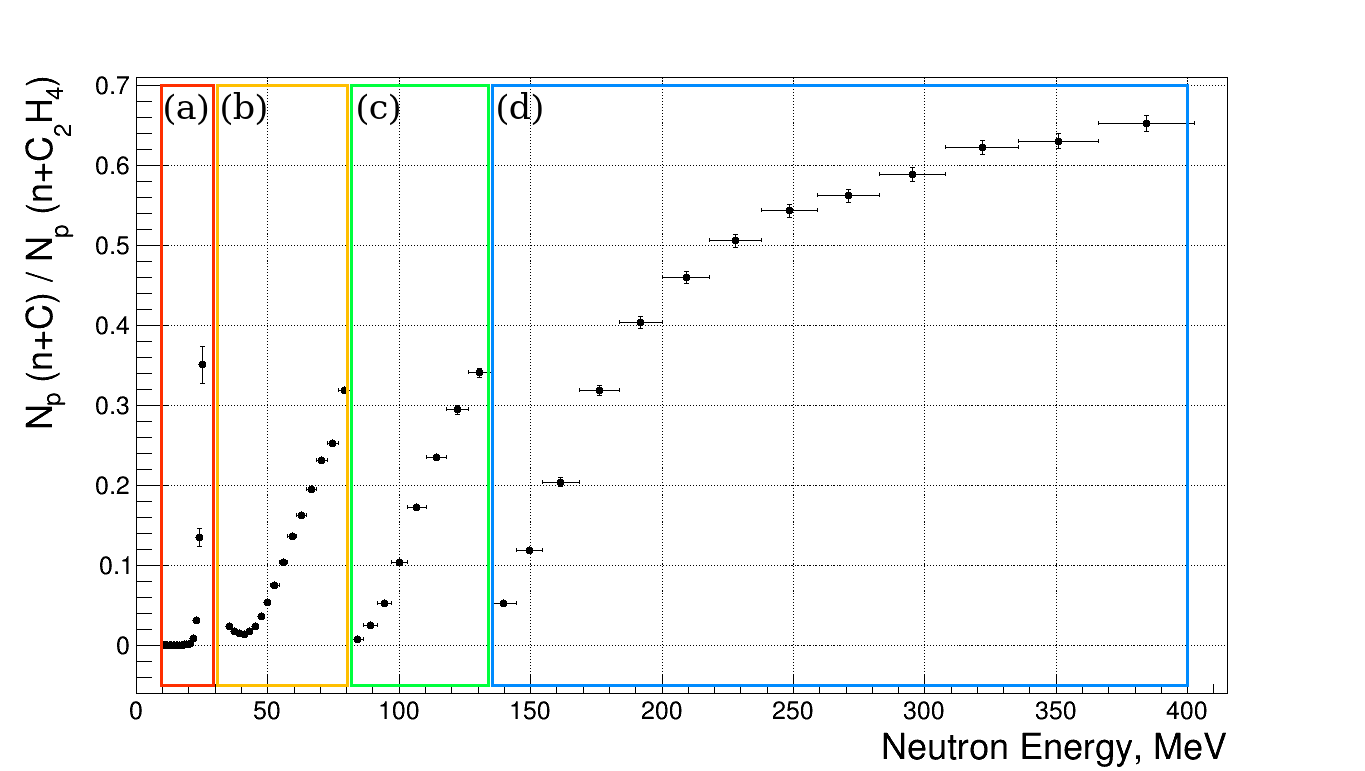}
\caption{Ratio between protons from n+C reactions, normalized to the number of carbon atoms contained in the polyethylene, and the total number of protons recorded when a C$_2$H$_4$ target is placed in the neutron beam. In the region (a), the events in coincidence between the two silicon detectors are shown. In regions (b), (c), and (d), the coincidences between the first two, three and four scintillators are shown, respectively. In each change of configuration, the number of events from the carbon shows a sharp drop, followed by gradual increase with energy.}
\label{fig:C_fraction}
\end{figure}

Figure~\ref{fig:C_fraction} shows the ratio between the number of protons produced by neutron induced reactions on carbon and the total number of protons coming from the target. The correlation was obtained after normalizing the number of counts produced by a carbon sample for the number of carbon atoms contained in the C$_2$H$_4$ target. In this plot, four regions are highlighted. In (a), the events in coincidence between the two silicon detectors are presented. The coincidences between the first two, the first three and all the plastic scintillators are in the (b), (c) and (d) regions, respectively. \\
It can be noted that the general trend of this graph is repeated for each region of coincidences: the fraction of background events is increasing as a function of energy and while exhibiting an abrupt reduction at each change of sub-detector combination. This behaviour is due to two factors. On one hand, as the energy increases the cross sections of the decay channels for $^{12}$C producing protons increase and, in addition more decay channels for $^{12}$C become open, therefore their contributions are more significant.
On the other hand, a reduction in the carbon background events is due to the multi-layer structure of the telescope that makes the detector more selective as more slabs are involved.

For both MS-RPTs, the efficiency was studied and background sources were characterised through Monte Carlo simulations.
The main difference between the two detectors is the extra background source present in the MS-RPTH. In fact, events coming from the first polyethylene sample are also detected. A full simulation of the entire experimental setup including the two polyethylene samples and the two telescope counters was thus developed in order to estimate the mutual background contributions. Both contributions were investigated. Back-scattered secondary particles from the second sample reaching MS-RPTL, and, on the other hand, particles from the first sample that can deposit significant energy in MS-RPTH~\cite{Terranova}. It was found that the former has no significant contribution over the whole energy range. On the other hand, MC simulations show protons from the first target reaching the second telescope are completely suppressed by imposing triple scintillator coincidence, a condition that corresponds to neutron energies above 100~MeV.

\subsection{RPT response to the neutron beam}
\label{sec:RPT_response}
\begin{figure}[ht]
\subfigure[Data - C$_2$H$_4$ sample]{\includegraphics[width=0.48\textwidth]{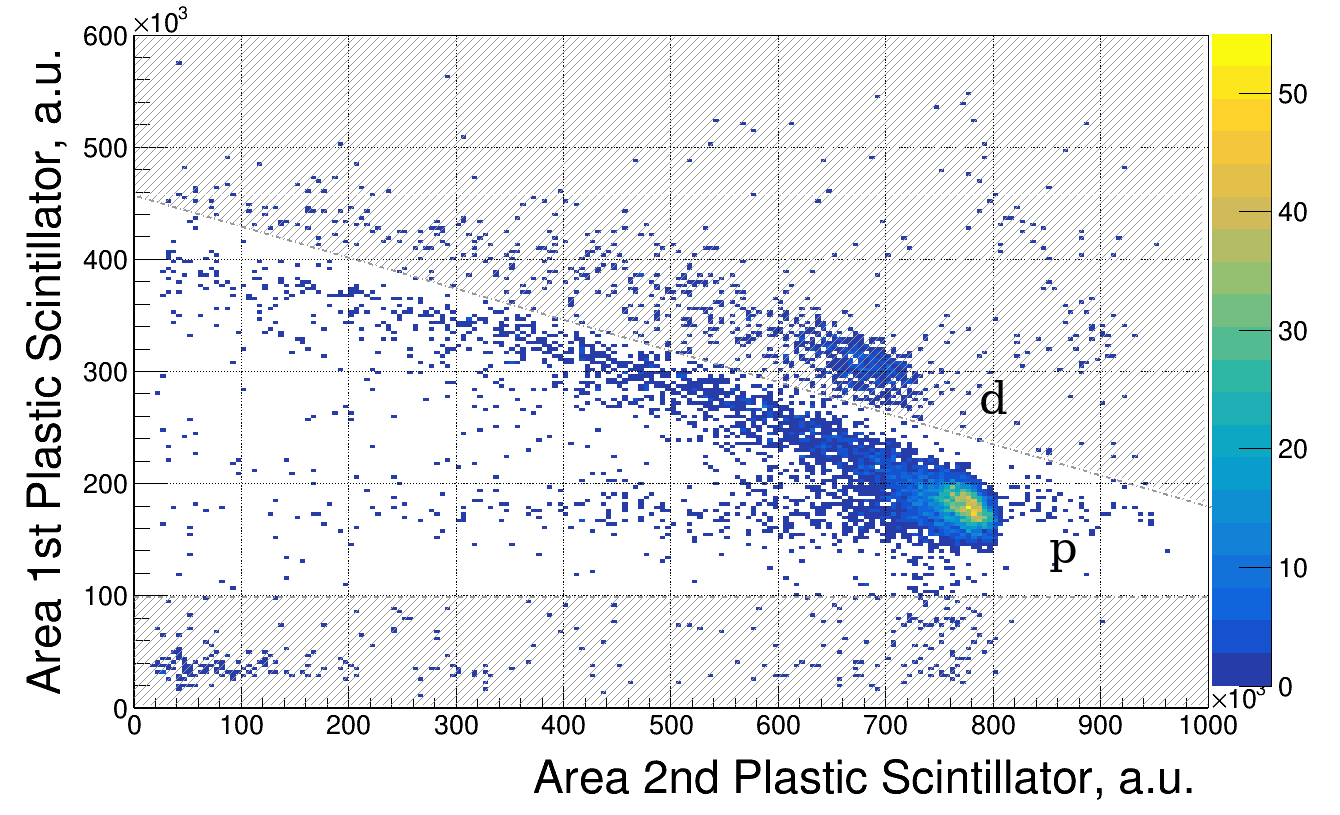}\label{fig:DEE_Poli_Dati}}
\subfigure[MC - C$_2$H$_4$ sample]{\includegraphics[width=0.48\textwidth]{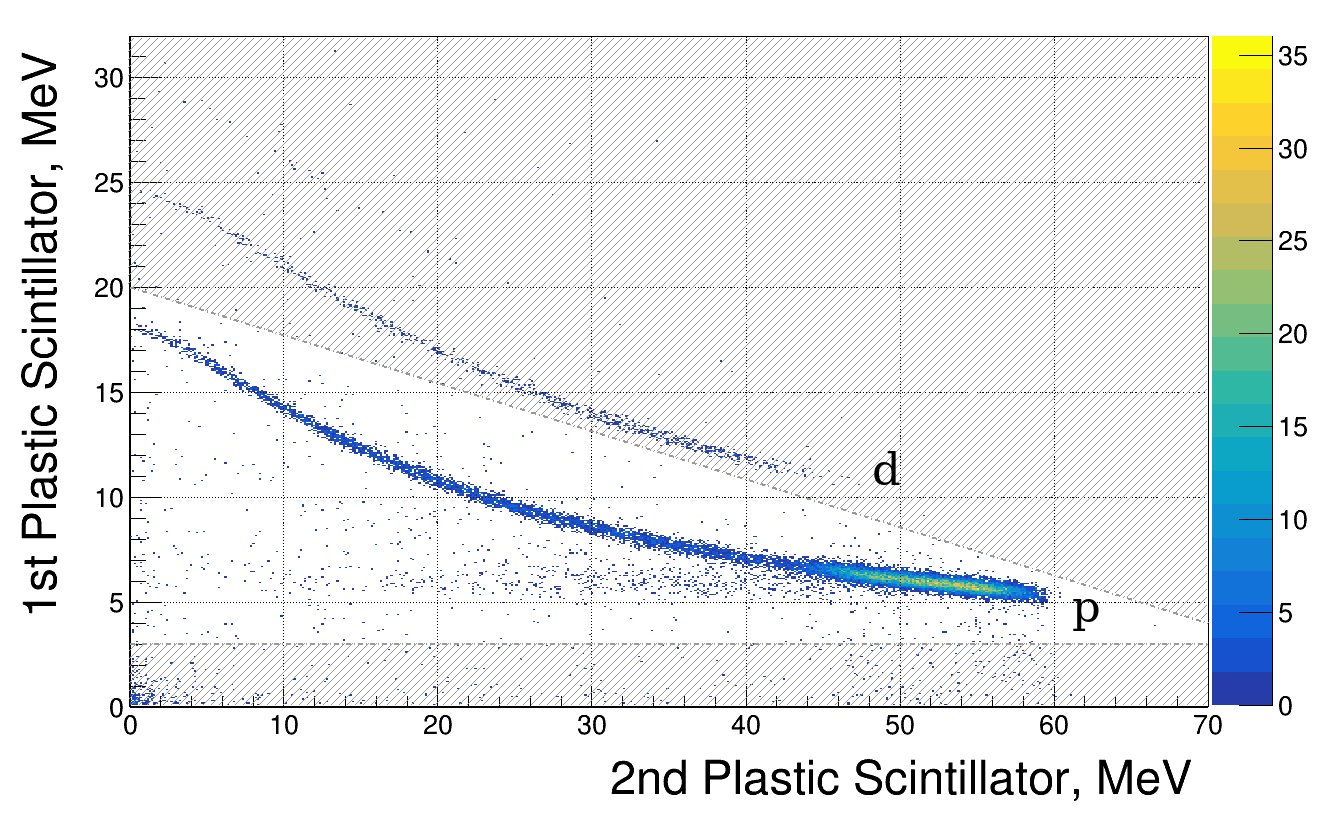}\label{fig:DEE_Poli_MC}}
\subfigure[Data - C sample]{\includegraphics[width=0.48\textwidth]{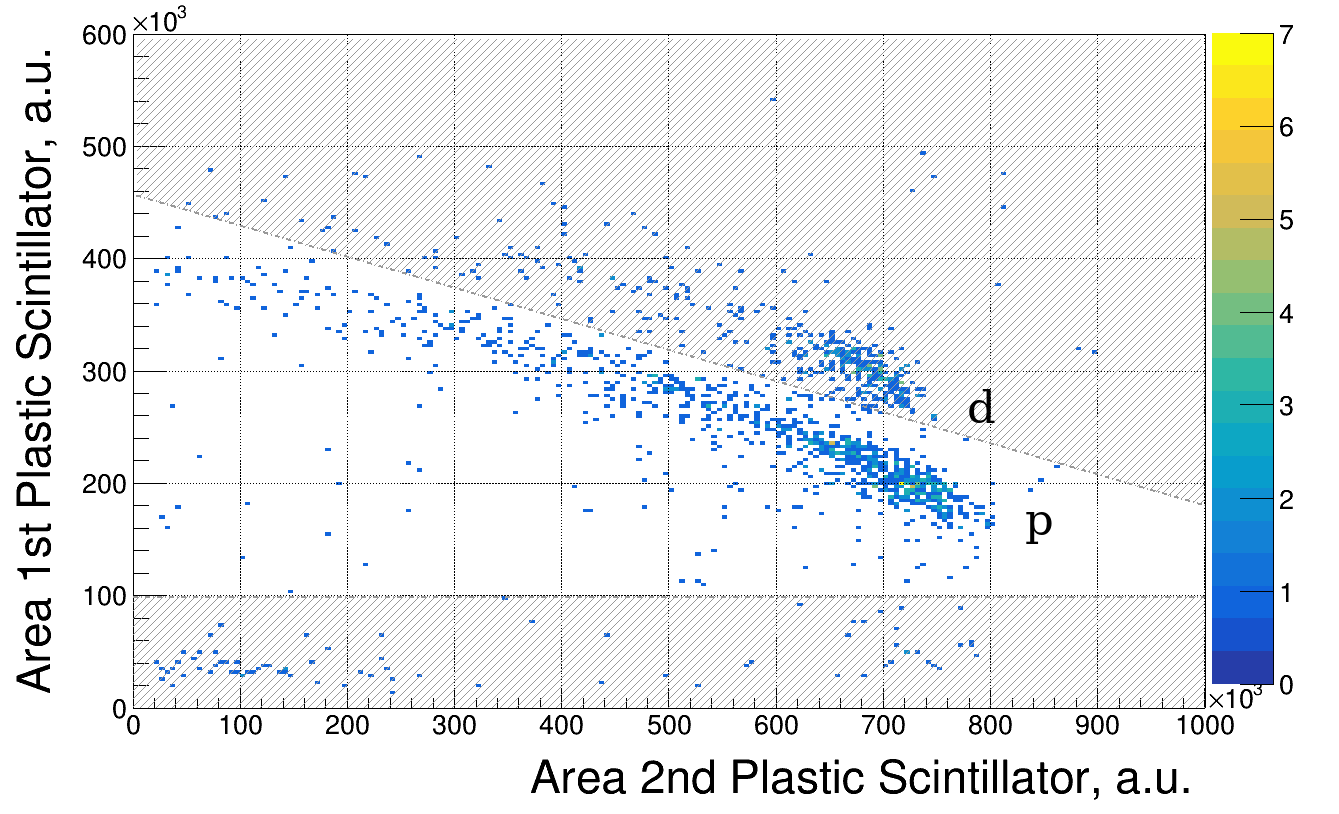}\label{fig:DEE_C_Dati}}
\subfigure[MC - C sample]{\includegraphics[width=0.48\textwidth]{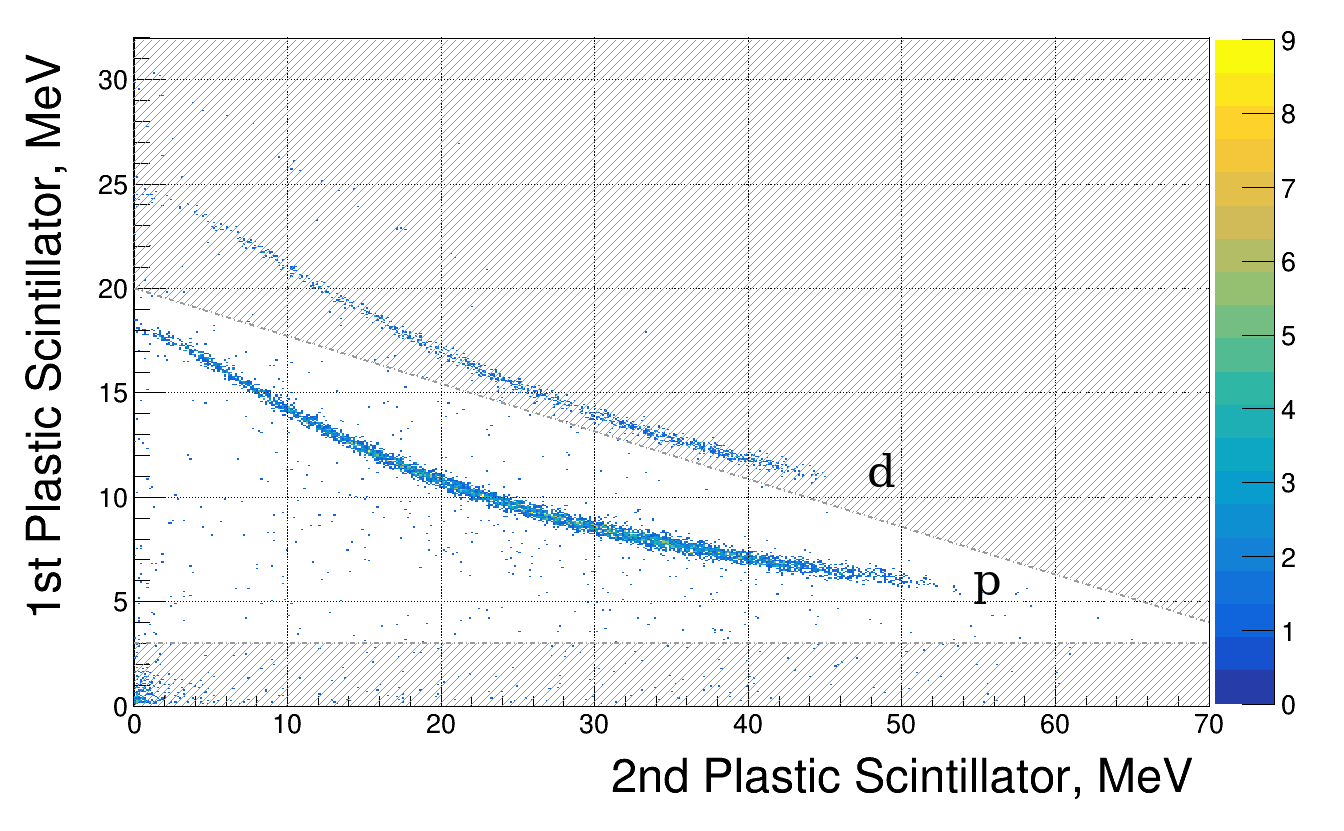}\label{fig:DEE_C_MC}}
\subfigure[Data - H]{\includegraphics[width=0.48\textwidth]{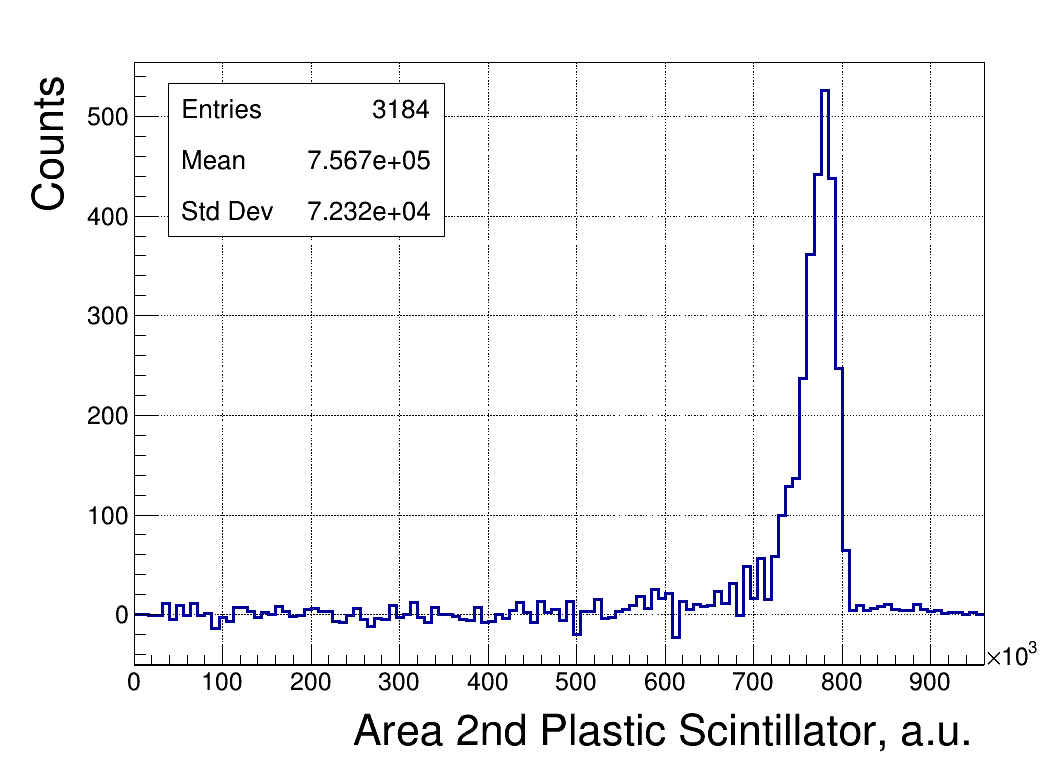}\label{fig:H_Dati}}
\subfigure[MC - H]{\includegraphics[width=0.48\textwidth]{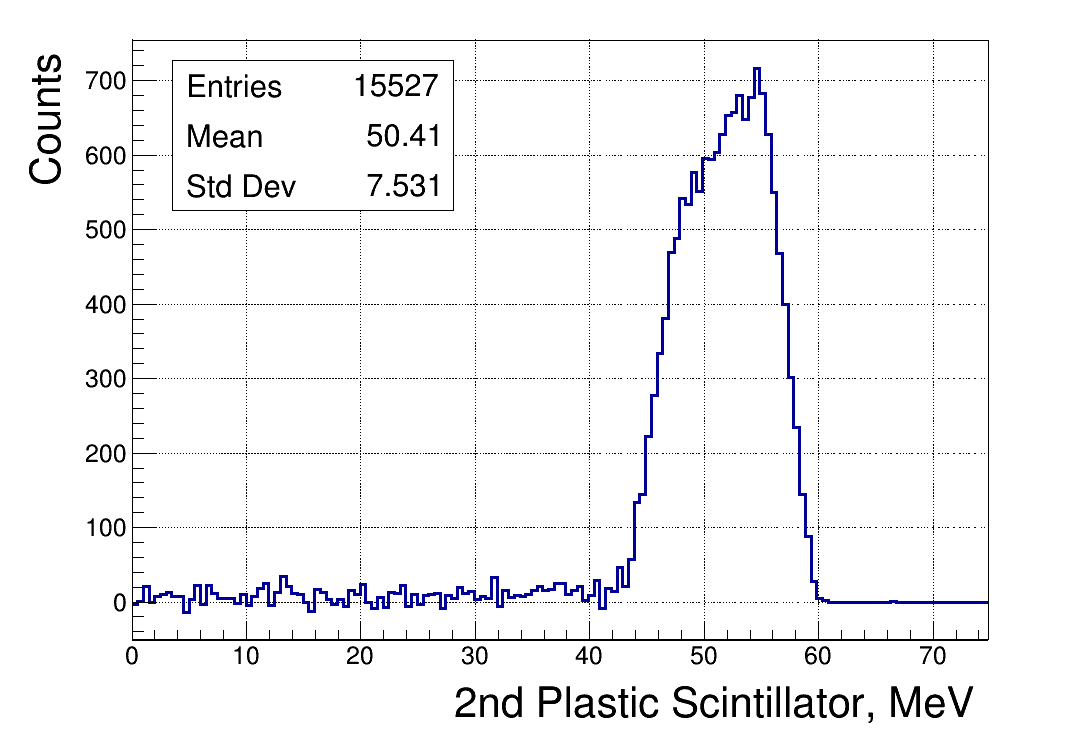}\label{fig:H_MC}}
\caption{Figures~\ref{fig:DEE_Poli_Dati} and~\ref{fig:DEE_C_Dati} display the $\Delta$E-E matrices produced by the experimental data, choosing the events of neutrons with energy of (74.8\,$\pm$\,2.2)~MeV hitting the 2~mm thick C$_2$H$_4$ and the 1~mm thick C sample, respectively. Figures~\ref{fig:DEE_Poli_MC} and~\ref{fig:DEE_C_MC} show the same $\Delta$E-E matrices but obtained through data from simulations. In each matrix, only the proton hyperbola was selected and the subtraction between the two samples was performed. The one-dimensional histograms in the bottom panels show the result of the subtraction between the two samples for the data, the figure~\ref{fig:H_Dati}, and for the simulations in~\ref{fig:H_MC}.}
\label{fig:DE_E_75MeV}
\end{figure}

Figure~\ref{fig:DE_E_75MeV} shows an example of experimental $\Delta$E-E matrices, taking into account the events produced by neutrons with energy of (74.8\,$\pm$\,2.2)~MeV, producing protons which stop in the second plastic scintillator. The plots~\ref{fig:DEE_Poli_Dati} and~\ref{fig:DEE_C_Dati} are based on experimental data using the 2~mm thick C$_2$H$_4$ and the 1~mm thick C target, respectively. 
The matrices~\ref{fig:DEE_Poli_MC} and~\ref{fig:DEE_C_MC} show the corresponding Monte Carlo simulation.
The two hyperbolas from protons and deuterons are clearly identifiable and events out of the gray zone were selected and counted. 
The subtraction between the polyethylene and the carbon, allowed us to isolate the contribution of n-p scattered protons from protons produced in the n+C reaction, obtaining the two one-dimensional histograms, shown in Figure~\ref{fig:H_Dati} for the experimental data and~\ref{fig:H_MC} for the MC simulation, respectively. While the behavior of the simulated data (left panels in~\ref{fig:DE_E_75MeV}) reproduces that of the experimental events (right panels in~\ref{fig:DE_E_75MeV}), the differences between the plots are due to the experimental energy resolution of the detector, not included in the MC simulation. 
\\

The discrimination between protons and deuterons in the $\Delta$E vs E plots is quite effective up to E$_n\,\approx\,$200~MeV deposited energy, as protons are already beyond the punch-through energy, whereas deuterons are still in the upper branch of the plot where the $\Delta$E is effective in the discrimination. Above that energy, the deuterons cannot be easily discriminated, but the subtraction of the data from the graphite samples basically removes all of them. 
In this case, the exclusion of background events occurs only through the subtraction of the normalized events coming from carbon from all events detected with the polyethylene target. \\
\begin{figure}[ht]
\centering
\includegraphics[width=0.95\textwidth]{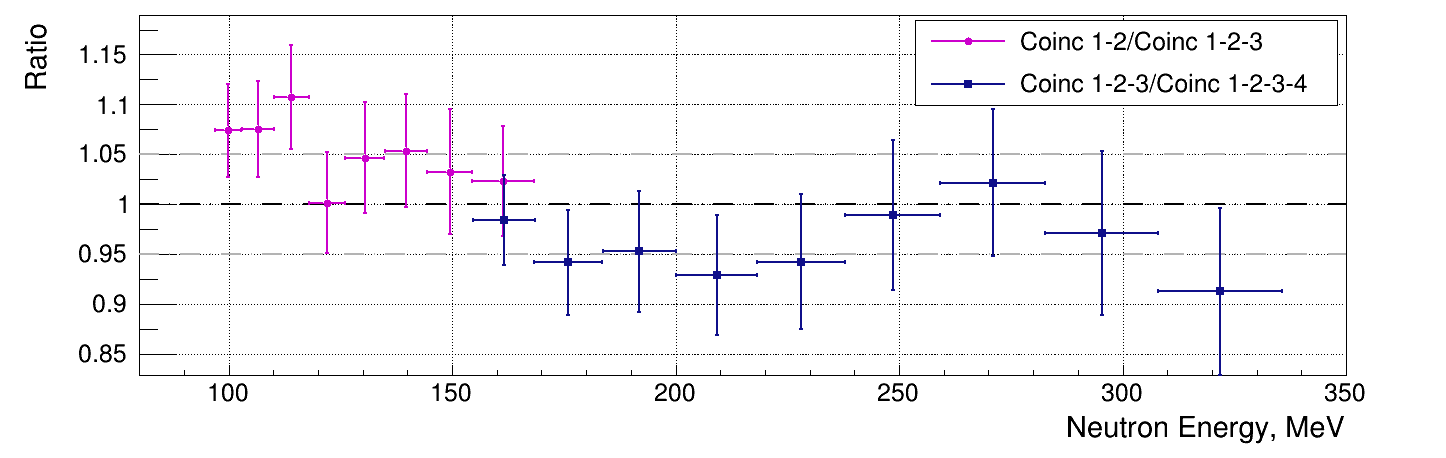}
\caption{The ratios between the different configurations, in the energy range where they overlap, are shown. Taking into account the coincidence between only the first two scintillators, in the energy range between 100 and 160~MeV, the telescope is working in the punch-through condition in fact the protons stop in the third scintillator. The same situation is for the energy range between 160 and 330~MeV requiring the coincidence 1-2-3 instead of coincidence among all the scintillators.}
\label{fig:Coinc_Anticoinc}
\end{figure}
Since this method is used for neutrons with energies from 200~MeV upwards, it is appropriate to find a way of testing the reliability of the results, especially when the protons traverse the telescope without stopping in it. Figure~\ref{fig:Coinc_Anticoinc} aims at demonstrating the goodness of the results when the telescope works in the latter configuration. From neutron energy of about 90~MeV, imposing only the coincidence between the first two scintillators, the flux can be extracted working in punch-through condition, while the coincidence among the first three scintillators is the right condition to require. A similar scenario occurs from about 160~MeV for the coincidence among the first three plastic scintillators. The ratios of the counts, already corrected for the dead-time, (see section~\ref{sec:deadtime}) confirm that, in the overlapping zones, the counts in the \textit{right configuration} of scintillators and in the one obtained with the previous couple of detectors are fully compatible within uncertainties. This validation allows to work with telescopes even in the high-energy region, where protons no longer stop inside the detector.

\subsection{Dead time correction}
\label{sec:deadtime}
\begin{figure}[ht]
\centering
\subfigure[]{\includegraphics[width=0.45\textwidth]{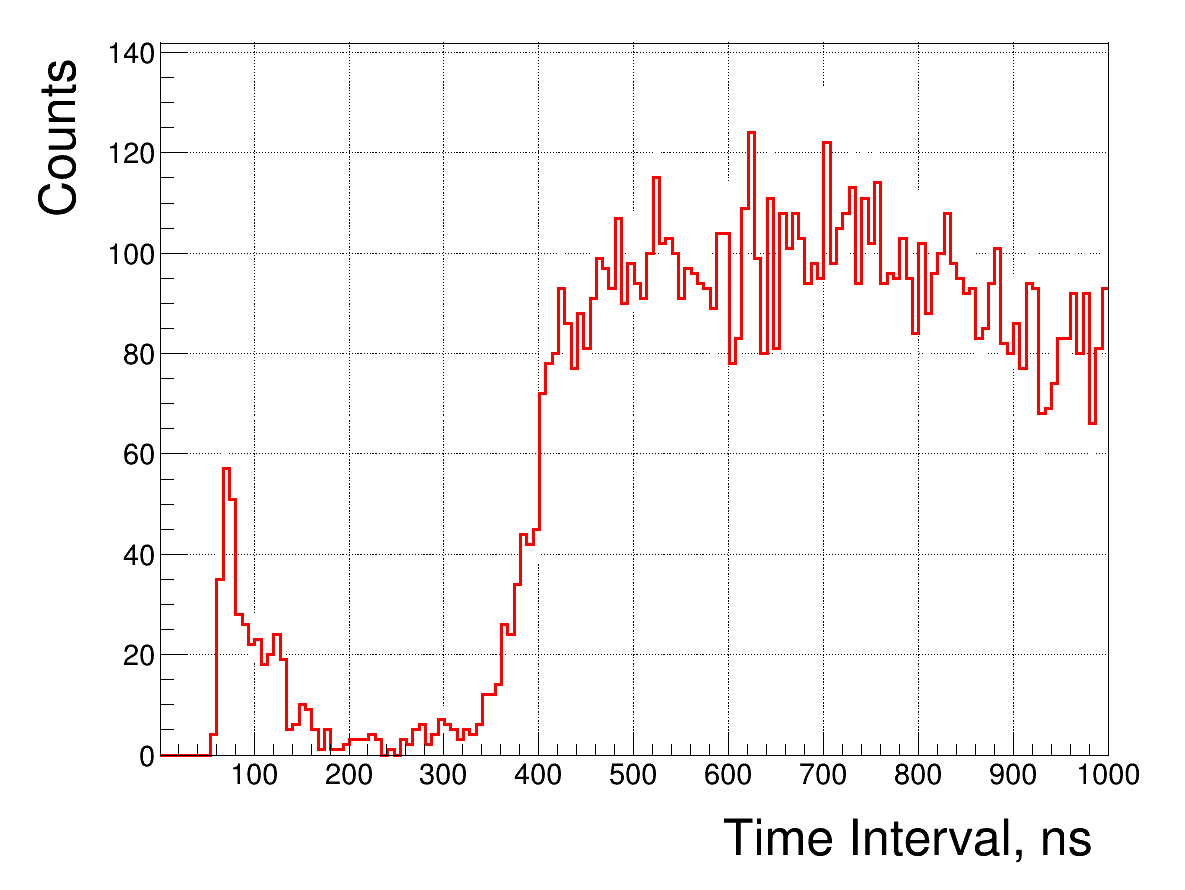}\label{fig:deadtime_sili}}
\subfigure[]{\includegraphics[width=0.45\textwidth]{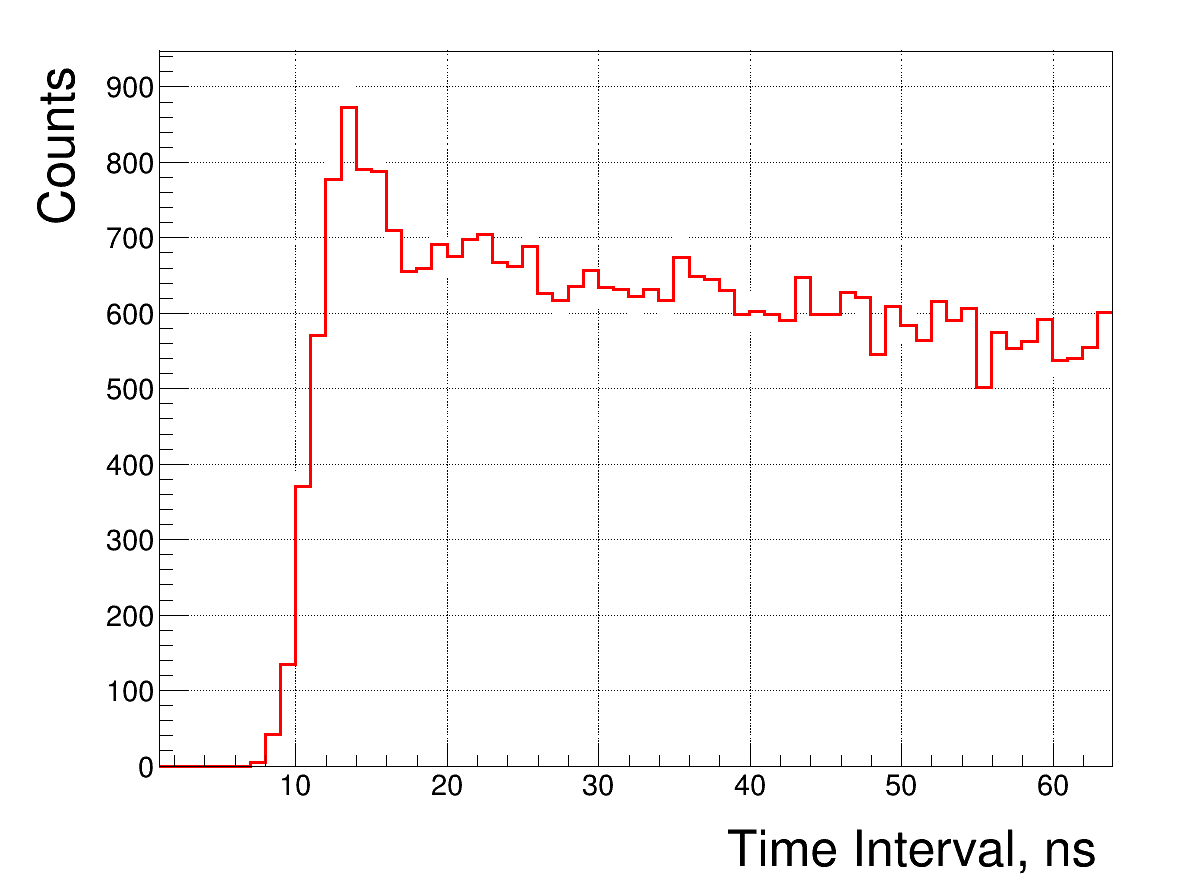}\label{fig:deadtime_sci}}
\caption{Time difference between two consecutive events recorded by a silicon detector, in figure~\ref{fig:deadtime_sili}, and a plastic telescope scintillator, in figure~\ref{fig:deadtime_sci}.}
\label{fig:deadtime_estimation}
\end{figure}

The time distribution between two consecutive events in the same detector was defined to estimate the typical dead time of each detector. Figures~\ref{fig:deadtime_estimation} shows an example of time difference distribution for the first silicon and one relative to the fourth scintillator, where a the dead time of about 400~ns for~\ref{fig:deadtime_sili} and 10~ns for~\ref{fig:deadtime_sci} can be estimated. 
In the analysis we used a  fixed dead-time of 450~ns and 13~ns and the detector's counting rates were corrected accordingly.
To estimate the uncertainty related to the dead time correction, we repeated the analysis with fixed dead time  ranging from 400 to 700~ns for the silicon detector and from 13 to 18~ns for the scintillator. In all cases, results are in agreement within 1\%.
The formula used for the correction, based on the approach studied by Whitten for time of flight facilities~\cite{Whitten}, was applied separately for dedicated and parasitic pulses. The real number of coincidences $N_{t,ev}(i)$ in the $i-th$ tof bin can be expressed as:
\begin{equation}
N_{t,ev}(i) \,=\, \alpha(i) \,  N_{t,0}(i)
\end{equation}
where $\alpha(i)$ is the tof-dependent dead-time correction and $N_{t,0}(i)$ is the number of events of coincidences recorded in the $i-th$ tof bin. \\
In the case of n consecutive sub-detectors in coincidence the correction factor can be calculated with the formula:
\begin{equation}
\alpha (i)\,=\, - \frac{N_{ppb}}{N_{t,0}(i)} \, ln \, \Biggl\{ 1- \frac{N_{t,0}(i)/N_{ppb}}{ \Bigl( \prod_{\phantom{i}d=1}^{\phantom{i}n} N_{d} \Bigr) \cdot  N_{t}} \Biggr\} 
\label{formula:deadtime}
\end{equation}
where $N_{ppb}$ is the number of protons per pulse, $N_d$ and $N_t$ can be defined as:
\begin{equation}
N_{d}\,= \Biggl( \sum_{k = i-\tau_d}^{i-1} 1- \frac{N_{d}(k) - N_{t,0}(k)}{N_{ppb}} \Biggr) \quad \textnormal{and} \quad N_{t} \,=\, \Biggl( \,  \sum_{k = i-\tau_d}^{i-1} 1- \frac{N_{t,0}(k)}{N_{ppb}} \, \Biggr)
\end{equation}
where $\tau_d$ is the dead time of a sub-detector, $N_d$ is the number of events for single sub-detector $d$, integrated from the tof bin $i-\tau_d$ up to $i-1$, subtracted for the total number of events in coincidence and $N_t$ is the integrated number of events in coincidence. In this way, the random contribution of individual sub-detectors can be disentangled from the true coincident events, thus avoiding an overestimation of the correction.

\begin{figure}[ht]
\centering
\subfigure[Silicon detectors]{\includegraphics[width=0.45\textwidth]{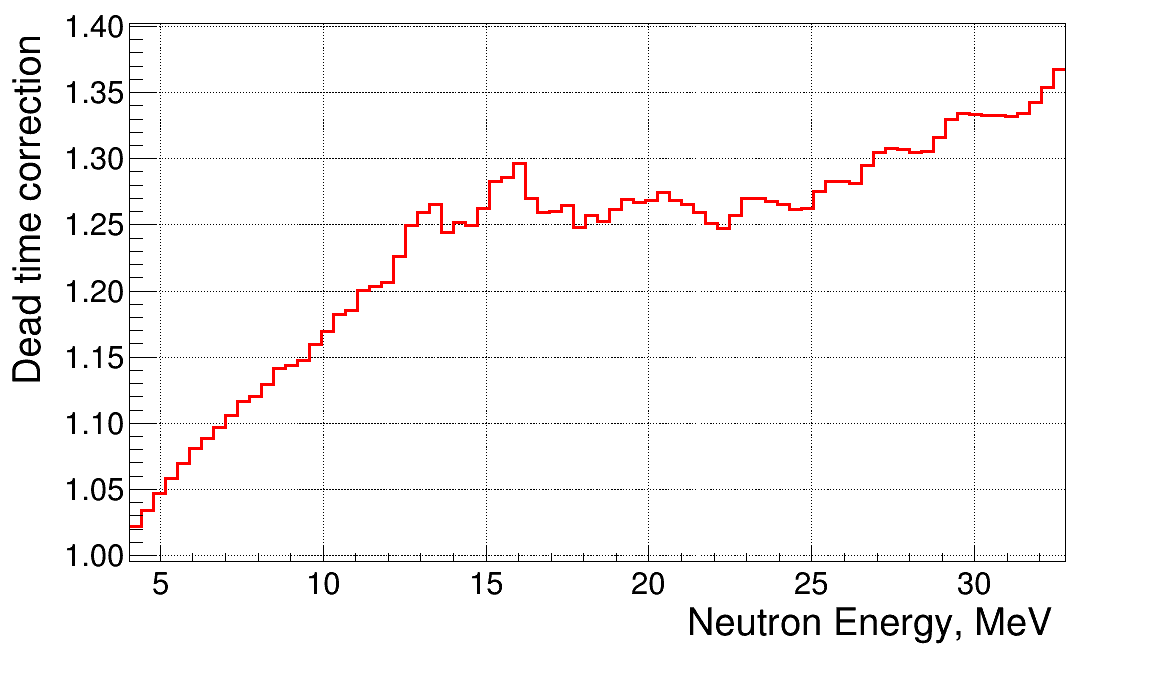}\label{fig:deadtime_sili_corr}}
\subfigure[Plastic scintillators]{\includegraphics[width=0.45\textwidth]{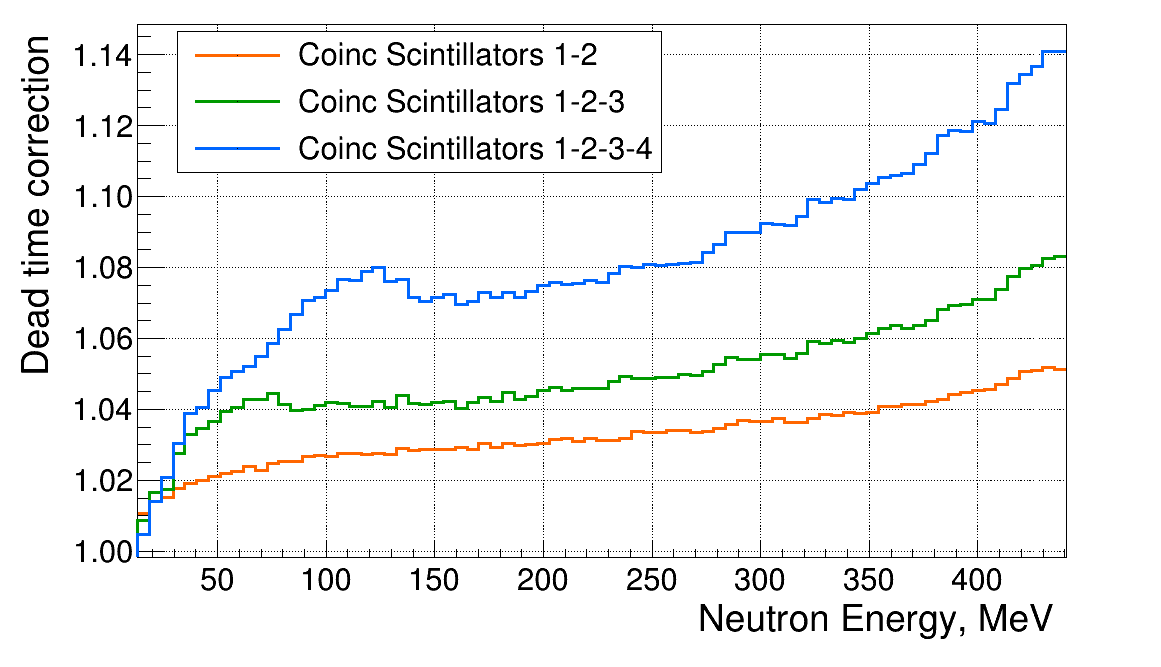}\label{fig:deadtime_sci_corr}}
\caption{Dead time correction calculated with the formula~\ref{formula:deadtime} for the coincidence between the silicon detectors (in parasitic pulse mode), in figure~\ref{fig:deadtime_sili_corr}, and for the different configurations between the plastic scintillators (in dedicated pulse mode), in figure~\ref{fig:deadtime_sci_corr}.}
\label{fig:deadtime_estimation_corr}
\end{figure}

\begin{figure}[ht]
\includegraphics[width=1.0\textwidth]{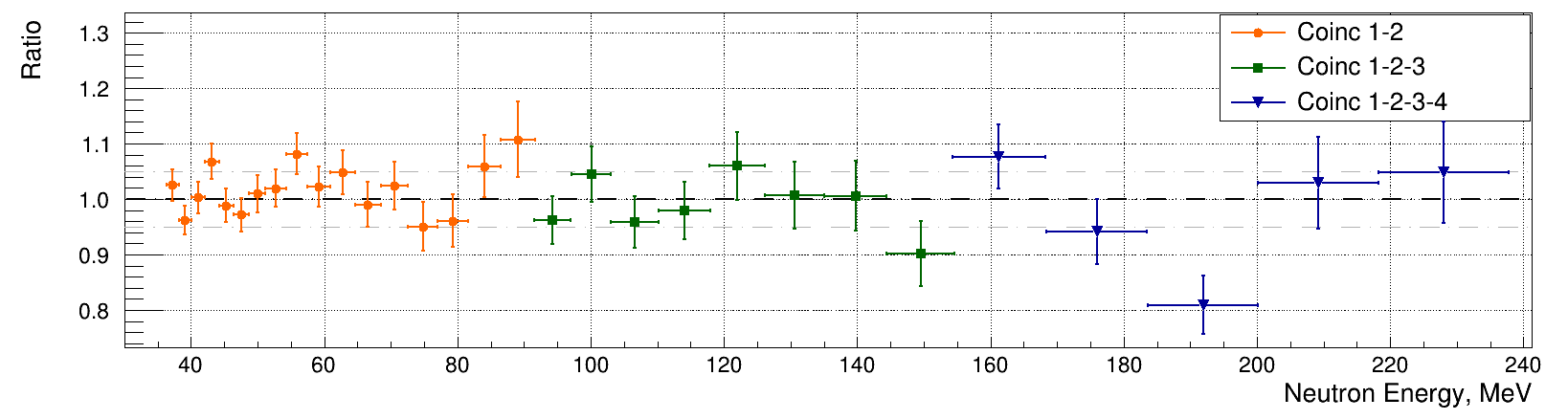}
\caption{The ratio between the counts recorded in dedicated and the parasitic mode of the PS pulse corrected for the dead time factor.}
\label{fig:Para_Dedi}
\end{figure}

The corrections calculated for the various coincidence configurations are shown in figure~\ref{fig:deadtime_estimation_corr}. In the left panel, the correction factor considering the coincidence between the two silicon detectors for parasitic pulses is displayed. The right panel shows the three factors for the three coincidence configurations among the plastic scintillators, for dedicated pulses. The corrections are as high as 35\% for silicons and  14\% for scintillators.\\
Since the dead time correction was too high, for silicons only parasitic bunches were considered. On the contrary, for the plastic scintillators the dead-time correction was implemented separately in the events recorded for dedicated and parasitic pulses. The consistency of the two results confirms the validity of the applied method, as shown in figure~\ref{fig:Para_Dedi} where the ratios between the dead time corrected counts in parasitic and dedicated pulses are reported. 

\subsection{Systematic uncertainty}
From the characterisation of the RPTs, it is possible to extract the sub-components involved in the assessment of the systematic uncertainty associated with the measurement and with the analysis techniques. Corrections were then worked out and applied.
To evaluate the uncertainties affecting the measurement of the neutron flux, in particular those related to the analysis of the detection procedure, it is appropriate to divide the neutron energy range measured by the MS-RPTL into three regions, diversified by the different detectors used or by the different working conditions: 
i) from 10 to 30~MeV the analysis is characterized by the coincidence between the two silicon detectors;
ii) from 38 to 200~MeV the protons stop inside the MS-RPTL; 
iii) for energies higher than 200~MeV, the protons exit from the last plastic scintillator and the telescope works in punch-through condition.
Regarding the MS-RPTH, working in the highest energy region from about 150~MeV, there are two regions, separated at 200~MeV. \\
In table~\ref{tab:unc_PRT} all the uncertainty components are listed, for the different energy ranges. \\
One of the uncertainty components is the attenuation of the neutron beam trough the fission detectors. This effect was estimated with MC simulations.  
About 0.8\% of the neutrons interacting with the PPFC escape the sensitive area of the uranium samples inside the PPACs. Since the contribution of PPACs in neutron reduction is negligible (less than 0.05\%), the neutrons reaching the polyethylene target are still 99.5\% of the starting beam, therefore 0.5\% contribution is included as upper limit in the calculation of the systematic uncertainties.
On the other hand, the effect on the beam due to the first C$_2$H$_4$ target is already considered in the study of the MS-RPTH efficiency.

\renewcommand\arraystretch{1.3}
 \begin{table*}[h]
 \centering
  \footnotesize
\caption{Summary of uncertainties related to the extraction of the neutron beam flux in the $^{235}$U(n,f) cross-section measurement.}
\centering
\begin{tabular}{lrrr} 
Source of & Uncertainty & Uncertainty & Uncertainty  \\
uncertainty  & E$_n$\,=\,[10-30]~MeV & E$_n$\,=\,[38-200]~MeV &  E$_n$ $>$ 200~MeV \\
\hline
\hline
C$_2$H$_4$ mass & 0.4\% & 0.2-0.6\% & 0.2-0.6\% \\
C mass & 0.9\% & 0.2-0.6\% & 0.2-0.6\% \\
Isotropic composition of PE & 1.5\% & 1.5\% & 1.5\% \\
Signal reconstruction & 1.8\% & 0.5\% & 0.7\% \\
Events selection in the $\Delta$E-E matrix & 5.0\% & 2.0\% & 2.0\% \\
Dead time correction &  2.0\% & 1.0\% & 1.0\% \\
Telescope position &  0.9\% & 1.1\% & 1.2\%  \\
Beam profile & 0.5\% & 0.5\% & 0.5\% \\
Beam transmission through PPFC, PPAC & 0.5\% & 0.5\% & 0.5\% \\
\hline
\end{tabular}
\label{tab:unc_PRT}
\end{table*}

\section{Parallel Plate Avalanche Counters}
\label{sec:ppac}
\begin{figure}[ht]
\centering
\subfigure[]{\includegraphics[width=0.35\textwidth]{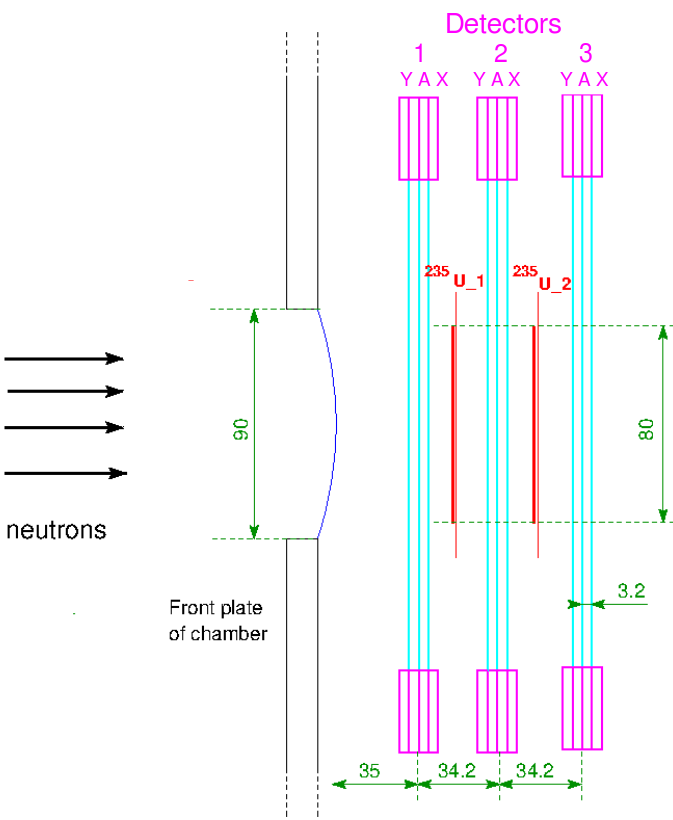}\label{fig:3riv}}
\subfigure[]{\includegraphics[width=0.53\textwidth]{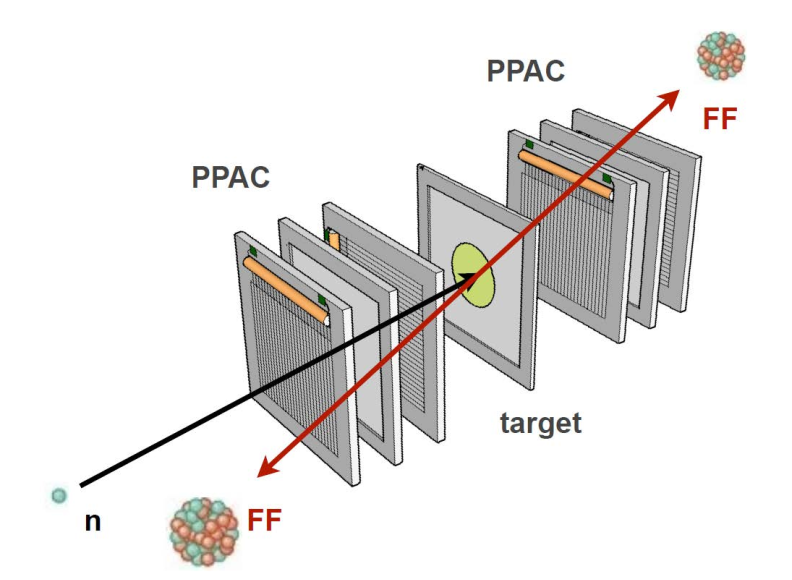}\label{fig:PPAC_coinc}}
\caption{(Left) Schematic view of the experimental setup with three PPACs and two uranium samples in between. All dimensions are expressed in mm. (Right) Fission detection concept using two PPACs surrounding the target. The fission fragments emitted from the target cross the two detectors.} 
\end{figure}

The Parallel Plate Avalanche Counters used in this work, developed at IPN-Orsay~\cite{Stephan, Tassan-Got}, are discussed in detail in ref.~\cite{Colonna, 2014NIMPA.743...79T}. Each PPAC consists of 3 parallel plate electrodes, a central anode surrounded by two cathodes, separated by a 3.2~mm gap, in order to maintain a high electric field and to reduce the time spread, thus leading to a good time resolution. 
The electrodes, with an area of 20×20 cm$^2$, are made of 1.7~$\mu$m thick mylar foils, coated with 700-nm thick gold layer.
The electrodes are enclosed in a container filled with octafluoropropane (C$_3$F$_8$) maintained at low-pressure (4~mbar) in a forced flow regime. The low gas pressure combined with the high constant electric field produced between the plates (600~V over 3.2~mm) create the conditions of a proportional regime.
The time properties of the output signal are related to the fast component of the signals, typically of the order of nanoseconds, ensuring a time resolution of 200~ps.
In the cathodes, the deposited gold is divided into 2~mm wide strips (with a width of 1.9~mm and a distance of 0.1~mm between two adjacent strips) connected to a delay line allowing the reconstruction of the fragment interaction position in the detector.
The delay line consists of a 20~cm of plastic rod with a coiled copper wire and an intermediate space of 6~mm at each side connects the delay line to the preamplifiers. 
The time difference between delay line outputs provides a one-dimensional position. Therefore, the combination of the signals from the two stripped orthogonal cathodes can provide the spatial information on the emitted fission fragment.
Clearly, this PPAC was conceived to detect in coincidence the two fission fragments emitted in the fission process. In fact, a detection cell is made of a uranium sample surrounded by two PPACs, as sketched in Figure~\ref{fig:PPAC_coinc}. It is important to mention that, the backing of the sample has to be thin enough to allow the backward-emitted fragment to be detected.
The fission reaction chamber, used in the measurement consisted of three PPACs with two $^{235}$U samples in between (Figure~\ref{fig:3riv}).

These detectors and the analysis procedure, which selects only events in coincidence between two consecutive PPACs within a time window of 10~ns, have already been used at n\_TOF to measure cross sections induced by neutrons with energy up to 1~GeV~\cite{Paradela}. Therefore, fission events selection from background events has already been established to operate with high precision. The only remaining factor is the accurate evaluation of the detector efficiency as a function of neutron energy in this specific configuration (described and studied in section~\ref{sec:PPAC_eff}).

\subsection{$^{235}$U Targets}
\begin{figure}[ht]
\centering
\subfigure[]{\includegraphics[width=0.45\textwidth, height=0.25\textheight]{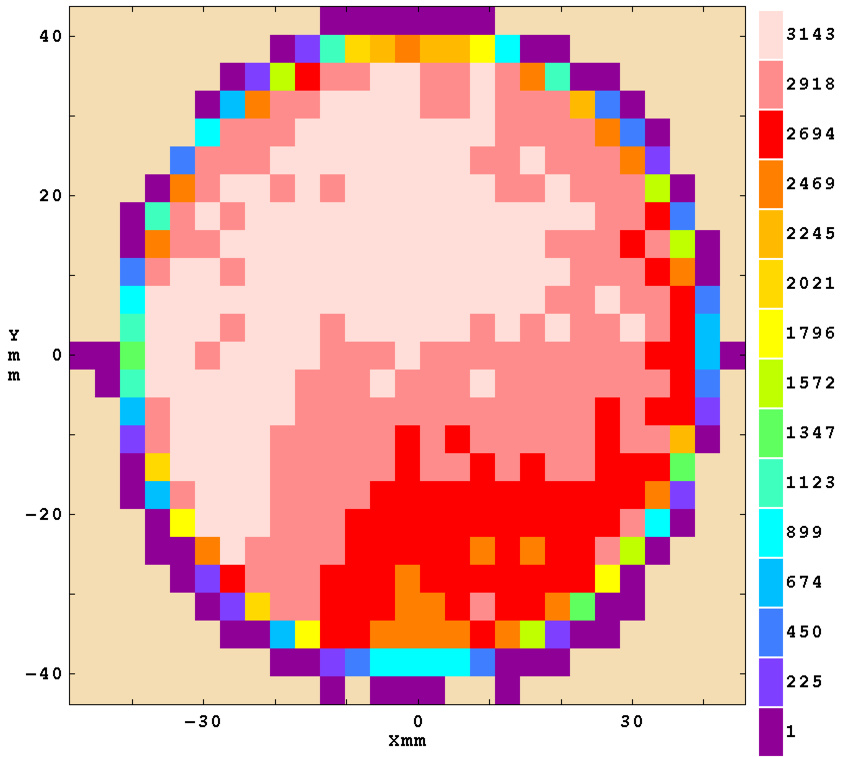}\label{fig:U5alpha}}
\subfigure[]{\includegraphics[width=0.45\textwidth, height=0.25\textheight]{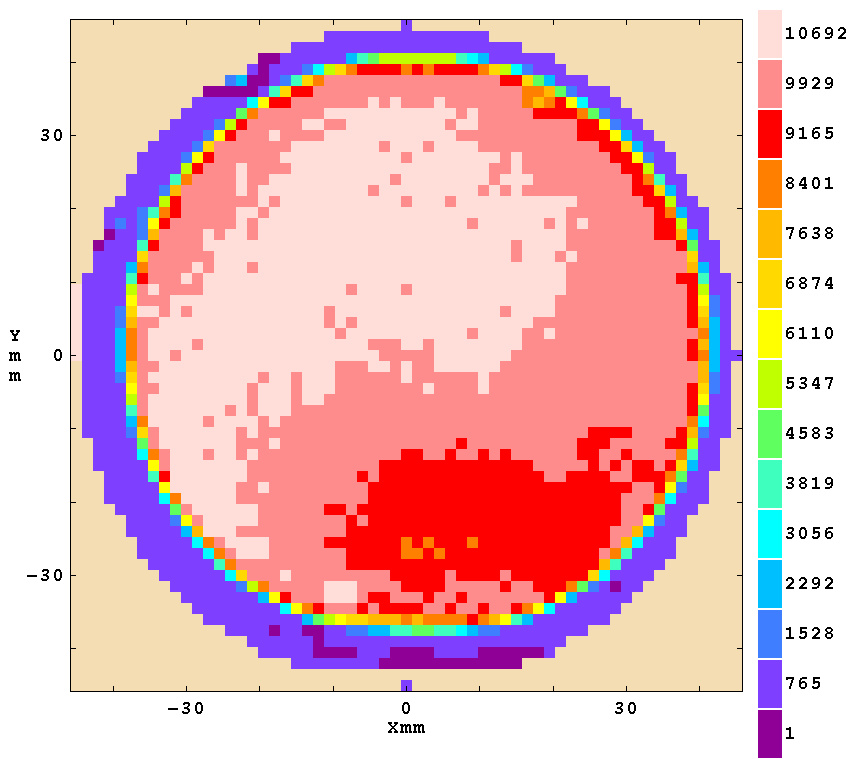}\label{fig:U5beam}}
\caption{Characterization of the thickness of the deposited material of one of the samples of $^{235}$U used in the measurement. (Left) $\alpha$ radioactivity from $^{234}$U isotope. (Right) Results of the Coulomb scattering measurement using a 2~MeV proton beam.} 
\label{fig:U5_Characterization}
\end{figure}

Two uranium targets, in total 27.11~mg of fissile material, were used in the PPACs detector system. The targets were produced by the radiochemistry group at the IPN d’Orsay using the molecular plating technique. They were made of a thin layer (around 0.3~mg/cm$^2$) deposited on a 80~mm diameter disk over 2~$\mu$m thick Al foil.
The aluminium foil was glued to a 1.5~mm thick epoxy frame with a 120~mm diameter centre hole in which the target had to be placed. 
From a mass spectrometry analysis the purity of the targets and the total number of nuclei was measured with an accuracy better than 1\%. The $^{235}$U samples have a purity of 92.699(5)\%, the remaining 7.300(5)\% is divided among isotopic impurities of $^{238}$U (6.283(6)\% in number of atoms), $^{234}$U (0.7472(15)\%) and $^{236}$U (0.2696(5)\%). The thickness of the radioactive targets is measured by counting the $\alpha$ radioactivity at 164~mm distance with a collimated silicon detector. In the case of $^{235}$U, the recorded events came mostly from the $^{234}$U contaminant. This technique has a spatial resolution of 3.5~mm (in x and y), and the result for one of the samples is shown in Figure~\ref{fig:U5alpha}. 
Furthermore, the uranium samples were exposed to a 2~MeV proton beam and the thickness was studied by exploiting the Coulomb scattering, with a step of 1.5~mm in the x and y position. The result of the analysis is reported in figure~\ref{fig:U5beam}.

The spatial distributions of the thickness of the uranium in the samples extracted through the two methods are compatible within the 1\%, taking into account the difference in spatial resolution.
Both measurements resulted in an average thickness for uranium samples of (0.280\,$\pm$\,0.003) mg/cm$^2$.

Recalling that each one of the 3 PPACs can locate the crossing point of the fission fragment in X and Y and assuming a back to back emission, the position of the emission point on the target was obtained. The accurate areal density of the region of the samples involved in the reactions can be calculated by the convolution of the beam profile and the sample characterization (figure~\ref{fig:U5_Characterization}).
Hence, where the $^{235}$U targets are hit by the neutron beam, the thickness of the sample is a factor (1.015\,$\pm$\,0.001) higher than the average value.

\subsection{PPAC detection efficiency evaluation}
\label{sec:PPAC_eff}
The main factor limiting the PPAC detection of fission fragments is the energy loss of fission fragments in the different layers of PPAC. In particular, the higher the angle, the longer the path through the successive layers of matter the fission fragment needs to go through. This introduces a limiting solid angle for the detection, which is more restrictive than the geometrical limitations. The consequence of the angular cut is a sensitivity to the anisotropy, particularly significant for energies above 1~MeV.
Since the coincidence between the two fission fragments is required to identify a fission event, the efficiency was derived accordingly, i.e. coincidences of both anode and cathode.

\begin{figure}[ht]
\centering
\subfigure[Target 1]{\includegraphics[width=0.45\textwidth]{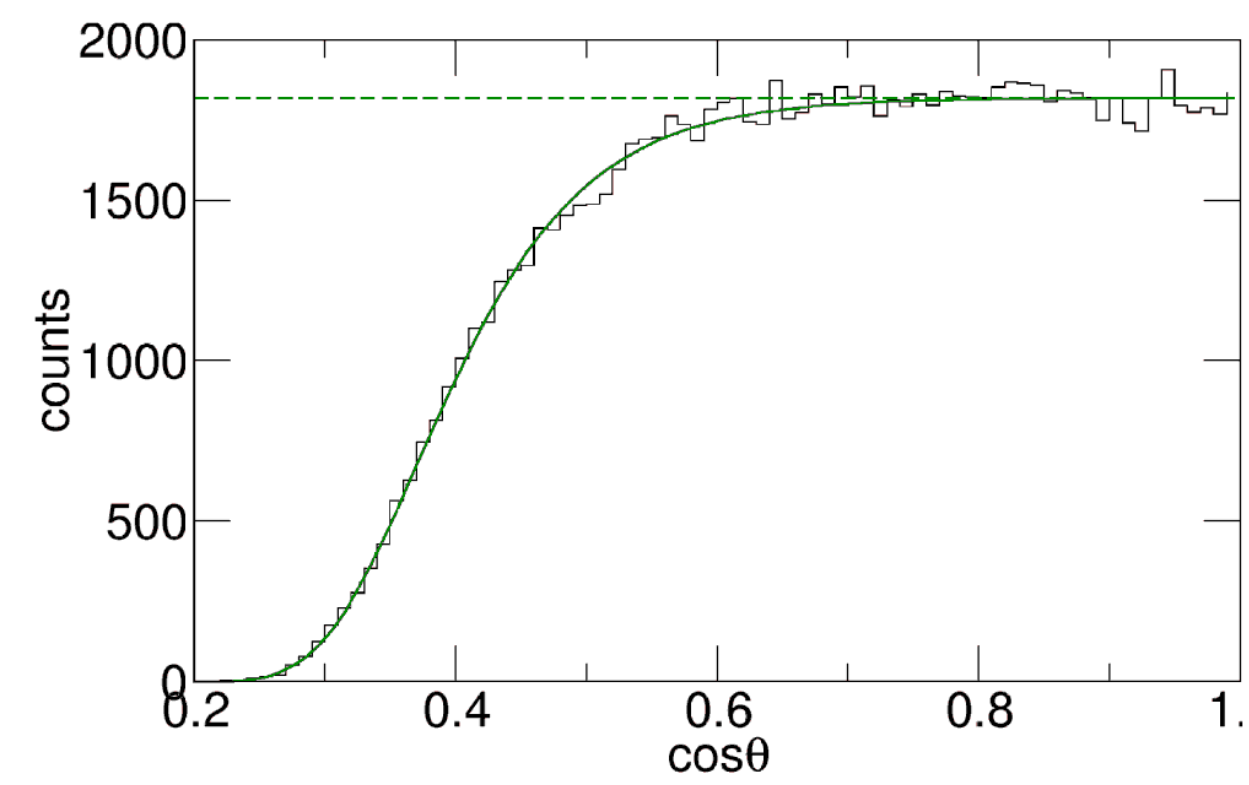}\label{fig:effi_1}}
\subfigure[Target 2]{\includegraphics[width=0.45\textwidth]{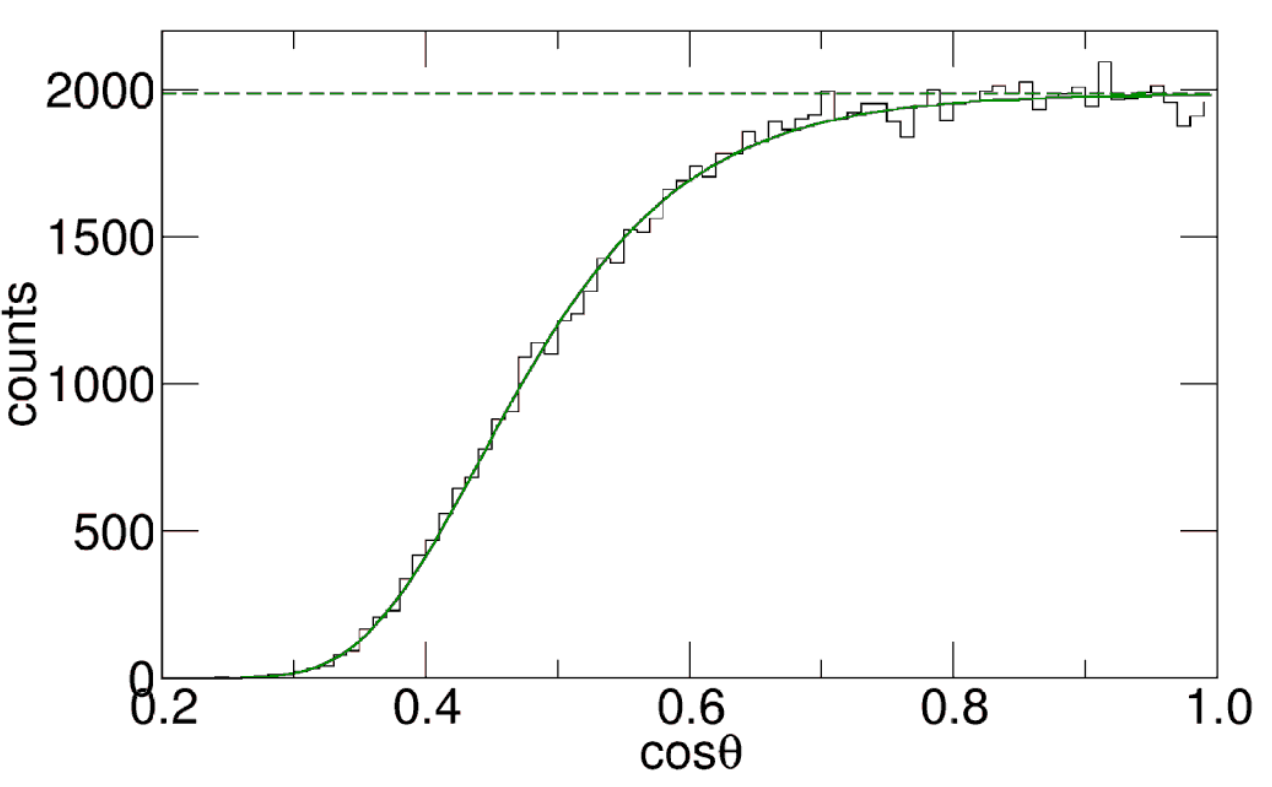}}
\caption{Distribution of the counts relative to cos\,$\theta$, obtained by the four localisation signals, of the two uranium targets, for low neutron energy (E$_n\,<$10~keV). In this energy region, the emission is isotropic, therefore,  a 100\% angular  efficiency would lead to a constant number of counts as depicted by the horizontal dashed green line.}
\label{fig:efficienza1_2}
\end{figure}

It is worth recalling that the forward-emitted fragment has to cross the aluminum sample backing of 2~$\mu$m, before reaching the PPAC active volume.
From the fission fragment trajectory and the measured angle $\theta$ (the angle between the beam axis and the segment connecting the two recorded fragments), it was possible to calculate the geometric efficiency of PPACs as a function of cos\,$\theta$. In fact the fission events triggered by low energy neutrons, e.g. E$_n$\,$<$\,10~keV, are emitted isotropically.
This is illustrated by Figure~\ref{fig:efficienza1_2}, which shows the cos\,$\theta$ spectrum obtained from the two $^{235}$U targets.

The main assumption here is that fission trajectories perpendicular to detectors and samples (therefore, parallel to the beam) are detected with an efficiency of 1 at all the energies. This is corroborated by the shape of the detected angular distribution which is flat in the vicinity of cos\,$\theta$=1. In figure~\ref{fig:efficienza1_2} the dashed green line shows the angular distribution if the efficiency was 1 at all angles.
The solid curves are fits of the efficiency with the formula:
\begin{equation}
\centering
\varepsilon (\cos\,\theta\,|\,p_0, p_1, p_2 ) = \frac{1}{\left( 1 + \exp\left( \frac{p_0 - cos\,\theta}{p_1}\right) \right)^{p_2}}
\end{equation}
where $p_0$, $p_1$ and $p_2$ are fitting parameters.

To determine the detection efficiency, it is necessary to normalize the plateau to 1 and integrate the distribution over cos\,$\theta$, which is equivalent to calculating the areas under the solid green lines. The efficiencies obtained from the coincidences between the first and the second PPAC, surrounding the first target, and the one from the second and the third PPAC, relative to the second uranium target, are different. The second target, as seen from the figures, is less efficient, due to the roughness of the deposit. The estimated values for the detection efficiency of fission events obtained from the analysis of Figure~\ref{fig:efficienza1_2} are 0.589 for target 1 and 0.510 for target 2, respectively. Simulations showed that the angle integration method is accurate to 1\% to compute the efficiency, however the statistical dispersion of the efficiency, in energy where it is expected to be constant, shows that the overall uncertainty is 2\%. 

\subsubsection{Angular distribution of emitted fragments}
\begin{figure}[ht]
\centering
\subfigure[Target 1 - 6.3~MeV $<$ $E_n$ $<$ 10~MeV]{\includegraphics[width=0.45\textwidth]{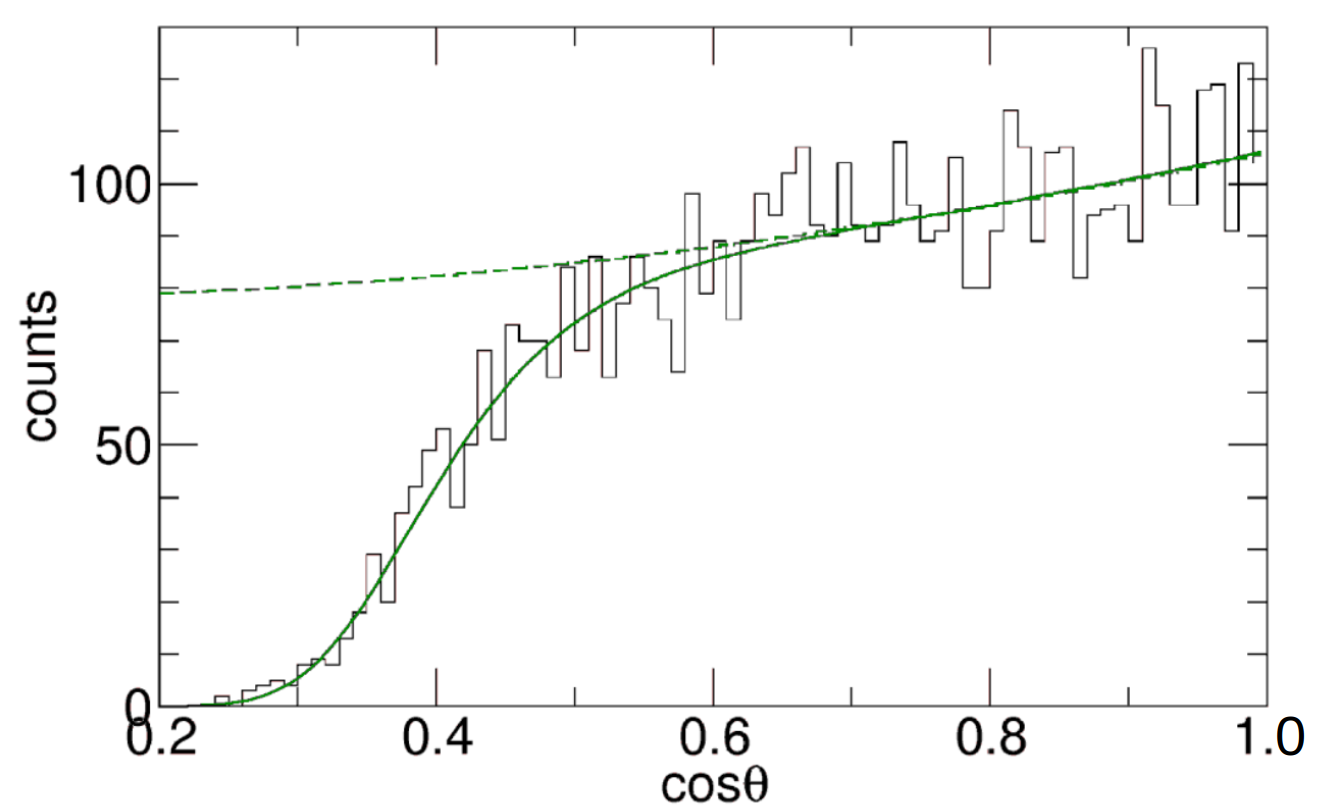}\label{fig:6.3_10}}
\subfigure[Target 1 - 630~MeV $<$ $E_n$ $<$ 1~GeV]{\includegraphics[width=0.45\textwidth]{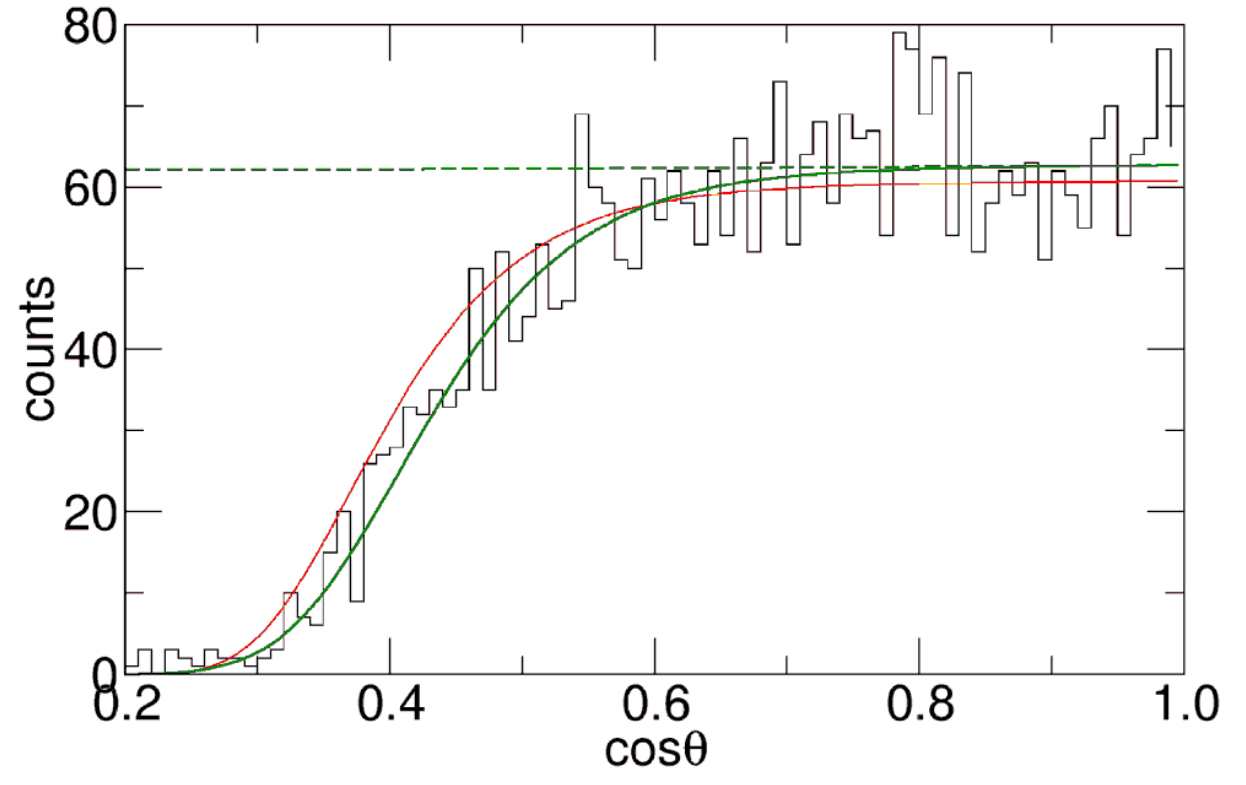}\label{fig:630_1}} 
\caption{Example of angular distribution for $^{235}$U, in two different neutron energy ranges. The green solid line is a result of the fit using the formula~\ref{fit}, the red one is the fit reported in figure~\ref{fig:effi_1}.}
\label{fig:anisotropy}
\end{figure}

The limiting angle implies a good efficiency at 0$^\circ$ and a complete insensitivity at 90$^\circ$. This behaviour makes the global detection efficiency dependent on the anisotropy of the fission fragment angular distribution, which in turn depends on the neutron energy.
The fission process anisotropy parameter is close to unity at low neutron energies, but undergoes particularly large variations at the multiple-chance fission thresholds and decreases steadily again, at intermediate energy. 
The anisotropy energy dependent parameter, used to correct for the angular distribution of the emitted fragments, is the one  measured by the Neutron Induced Fission Fragment Tracking Experiment (NIFFTE)~\cite{Geppert-Kleinrath} collaboration. 
The fission fragment angular distribution can be parameterized by a sum of even Legendre polynomials $P_L (cos \theta)$ series, for each neutron energy interval. Only the even Legendre polynomials are included in the sum, because of the backward–forward symmetry of the emitted fragments. 
The truncation in the sum over the order of the polynomial, $L$, is defined by the total angular momentum ~\cite{Vandenbosch}. 
However, the mean of the angular structures brought about by the averaging over the orientations of the entrance spin (J\,=\,7/2 for $^{235}$U), helps to reduce the number of $P_L$ to be included.
Indeed, Kleinrath and collaborators~\cite{Geppert-Kleinrath} showed that it is possible to stop at the second order for representing the angular distribution.
In particular, they demonstrated that by including higher-order terms in the calculation of the anisotropy parameter, the goodness of the fit parameter $\chi^2_{\nu}$ improves marginally, and the statistical uncertainty of the fit increases~\cite{Geppert-Kleinrath}.
Hence, the angular distribution can be parameterized as:
\begin{equation}
W (cos\,\theta_{cm}) = \sum_{L=2L even}^{L_{max}} a_L P_L (cos\,\theta_{cm}) \approx  a_0 + a_2 P_2 (cos\,\theta_{cm})
\label{formula:W}
\end{equation}
where $a_L$ is the energy-dependent coefficient of the L-th order Legendre polynomial obtained by fitting the experimental cosine distribution. From the condition of normalization of angular distribution:
\begin{equation}
\int_0^1 W(cos\,\theta_{cm}) d(cos \theta_{cm}) = 1 
\end{equation}
the value of the first parameter is obtained, $a_0$\,=\,1.\\
Therefore, the anisotropy parameter (A) can be written as: 
\begin{equation}
A \, = \, \frac{a_0 + a_2}{a_0 - a_2/2},
\label{formula:A}
\end{equation}
and from this formula it is possible to calculate the value of $a_2$, when the anisotropy is known.
Therefore, for each energy bin of the experimental data:
\begin{equation}
C(cos\,\theta)\,=\,(1\,+\,a_2\,P_2\,(cos\,\theta)) \, \varepsilon(cos\,\theta\,|\,p_0, p_1, p_2) \cdot p_3
\label{fit}
\end{equation}
where the parameters $p_0$, $p_1$, $p_2$ and $p_3$ are adjusted on experimental angular distributions. \\
Figure~\ref{fig:6.3_10} shows the result of the procedure for the first target, in the energy range from 6.3~MeV to 10~MeV, where the anisotropy is maximal. In this energy region, the data show a slope of the plateau due to the forward-backward peaking of the physical angular distribution. The green solid line is the fit performed and the dashed line shows the trend in case of efficiency equal to 1 at each angle. Instead, the distribution corresponding to the highest neutron energies, between 630~MeV and 1~GeV, is isotropic again, as shown in Figure~\ref{fig:630_1}. The green solid line is the result of the fit with formula~\ref{fit}, the red line, instead, is a fit using the low-energy parameterized efficiency $\varepsilon$ (corresponding to the green curve in figure~\ref{fig:effi_1}).

The actual efficiency has a lower cut-off angle with respect to the low-energy shape of the efficiency
To understand the kinematical effects and to control the accuracy of the angle reconstruction, for energies above 100~MeV we ran simulations with INCL4~\cite{2013PhRvC87a4606B} coupled to the deexcitation code Abla07~\cite{INDC2008}, and below 14~MeV we used the ENDF/B-VIII evaluation~\cite{2018NDS148_1B}. In both cases the slowing down of fission fragments was computed using SRIM~\cite{1985this.book.93Z, 2010NIMPB.268.1818Z, website_srim}.
In the simulation the detected angle was reconstructed as in the experiment, taking into account the angular and energy stragglings.
In the simulation the angular distribution is always isotropic, and we checked at 100~MeV and 1~GeV that the computed average linear momentum transfer is in accordance with the NIFFTE results~\cite{Hensle} and  with~\cite{Fatyga:1985xsy}, and allows us to test the boost effects.

\begin{figure}[ht]
\centering
\includegraphics[width=0.7\textwidth]{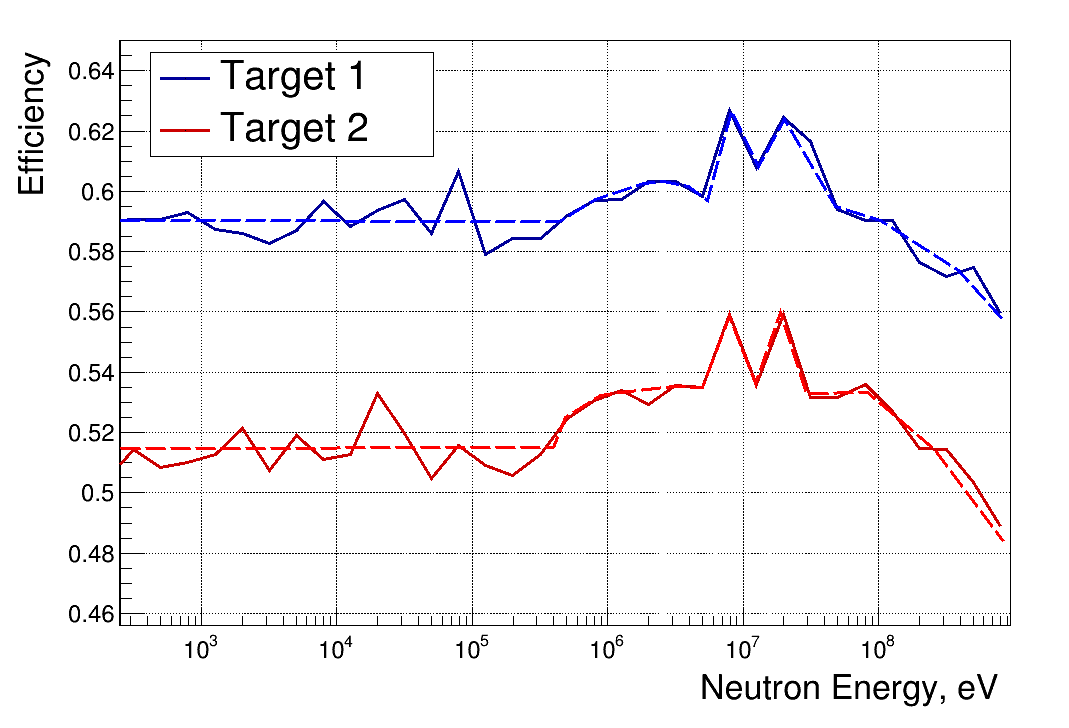}
\caption{Global efficiency calculated taking into account the geometric factor, the fission fragments absorption effect and the angular distribution of the products.}
\label{fig:2target_effi}
\end{figure}

First we checked the kinematical effects at high energy. The difference between the efficiencies close to 1~GeV and below 10~keV, although isotropy applies in both cases, is well reproduced. When looking into more details we see that the reduction of the efficiency at high energy results from a partial compensation between the boost effect which increases slightly the  efficiency and the more effective slowing down of fission fragments having lost many neutrons by pre- or post-evaporation. All in all the drop of efficiency above 50~MeV, depicted in figure~\ref{fig:2target_effi}, is quantitatively reproduced by the simulations. The second point we checked is the robustness of the efficiency  method based on the integration of the angular distribution when the boost tilts the fragment angles. In the simulation we can test the concordance between this value and the efficiency obtained by direct counting in the detectors. Below 50~MeV they always agree by better than 0.2\%. Above 100~MeV the angular method overestimates the "true" efficiency by a factor increasing with energy, reaching 2.8\% at 1~GeV, reflecting a boost effect. We have included this correction and adopted a linear correction in log(E) between 50~MeV and 1~GeV.

In summary the total detection efficiency $\eta$ is composed of the geometric acceptance, the in-medium fission fragment absorption and the effect due to the angular distribution of the emitted fragments. For each energy interval, $\eta$ was calculated according to:
\begin{equation}
\eta = \int_0^1 (1\,+\,P_2 (cos \theta)) \, \varepsilon(cos \theta \, | \, p_0, p_1, p_2) \, d(cos \theta).
\end{equation}
The resulting $\eta$ for the two targets are displayed in figure~\ref{fig:2target_effi}. The solid lines (blue and red) are constructed by taking 5 bin per decade, the dashed ones are the smoothed version, maintaining the general tendency and eliminating fluctuations due to statistics.

The global trend is common to both targets and the main features are:
\begin{itemize}
\item constant efficiency below 100~keV; in fact, in this energy range there are no effects due to anisotropy nor to the kinematics of the reaction;
\item increase of the efficiency around 10~MeV due to the physical angular distribution, and in particular the peaking at the opening of 2nd and 3rd chance fission;
\item drop beyond 100~MeV due to the the mass loss of fission fragments when excitation rises up, thereby increasing their slowing down. 
\end{itemize}
For each target an individual efficiency correction has been applied to the recorded counts. 
This is necessary because, although the two targets are subject to the same phenomena, the weights of the various factors are not the same, therefore the total efficiency is not equal for the two targets.

\subsection{Systematic uncertainty}
The constraint of coincidences between two adjacent detectors, on which the analysis of PPACs is based, ensures an almost background-free selection of fission events. However, a few specific steps require careful evaluation of the uncertainties involved in the analysis procedure.

Table~\ref{tab:unc} lists all uncertainties associated with the analysis of fission counts.
The range in the detection efficiency fit takes into account the effect of
the boost above 50~MeV. The uncertainty goes up linearly in log(E) up to 450~MeV. The range in the anisotropy correction uncertainty follows the range in anisotropy and scales accordingly. 

\begin{center}
\renewcommand\arraystretch{1.3}
 \begin{table}[h]
   \footnotesize
\caption{Summary of the uncertainties in counting FF events in the $^{235}$U(n,f) cross section measurement.}
\centering
\begin{tabular}{lr} 
& \\
Source of uncertainty & Uncertainty \\
\hline
\hline
Sample mass & 1.0\% \\
Reconstruction of FF trajectories & 0.4\% \\
FF detection efficiency  & 2.0 - 2.7\% \\
Correction for FF emission anisotropy & 0 - 1.2\% \\
\hline
\end{tabular}
\label{tab:unc}
\end{table}
\end{center}

\section{Conclusion}
A detection system was developed and characterised to measure the neutron-induced fission cross-section of $^{235}$U as a function of the neutron kinetic energy at the n\_TOF facility at CERN, in the 185-m long flight-path station.
On average, 10$^5$ neutrons per bunch are produced with energies between 10~MeV and 1~GeV, corresponding in EAR-1 to time of flights ranging from 100~ns to 2.5~$\mu$s after the $\gamma$-flash.
As a consequence, a common feature of the chosen detectors is a good timing resolution, so to cope with the narrow time-of-flight window of interest. In addition, a redundant measurement setup was adopted, with the aim of benchmarking the results while reducing systematic uncertainties related to the detection efficiency.
The experimental apparatus consisted of three flux detectors (RPTs) and two fission detectors (PPAC detectors and PPFC), thus allowing us to simultaneously record the number of neutrons impinging on the $^{235}$U samples and of fission events, as a function of the neutron energy.
Because of their correlation with the measurement of the final cross section, the comparison between the results obtained by the PPACs and a PPFC together with the extracted flux through the recoil proton telescopes will be shown in two additional articles (in preparation).

The measurement of the flux required an extensive background measurement and characterization to estimate the impact of the reactions induced by neutrons on the carbon contained in the polyethylene target. 
The coincidence technique, necessary to exploit the $\Delta$E-E technique and to perform particle identification, was efficiently applied thanks to the excellent timing properties of the detectors: the distribution of time difference for the events in coincidence between the two silicon detectors resulted to be within 50~ns
(FWHM), and for the plastic scintillators 1~ns (FWHM). 
The background induced by carbon reactions increases with the incident-neutron energy, and it ranges from a few percent up to a maximum of 60\% of the total proton events at 450~MeV.
Therefore, an extensive study with Monte Carlo simulations was carried out for a complete characterisation of the RPTs performances needed for the extraction of the neutron flux. \\
PPAC detectors have already been used in the n\_TOF facility and they have already been proven capable of detecting fission fragments generated by neutrons of energy up to 1~GeV.
In the present investigation, the correction for the angular distribution of the fission fragments, which is strongly anisotropic for neutron energy from a few MeV up to 200 MeV and for the Lorentz boost, has been thoroughly investigated. Since PPACs are limited in the angle of detection, this correction factor for this detector can reach up to a maximum of 6\% in our experiment.

The features and performance of the counter telescopes and the PPAC detectors described here strongly support the reliability of the resulting study of the $^{235}$U(n,f) cross section between 10 and 450~MeV. 

\acknowledgments

The authors would like to thank Davide Mancusi for his support provided on the validation of the PPAC detection efficiency distribution as a function of energy, through INCL and Abla simulations.
They wish to thank the n\_TOF local team for the assistance provided during and after
the experimental campaign, especially M. Bacak, and D. Macina.
The support provided by the PTB Working Groups ‘Solid State Density’ and ‘Scientific810 Instrumentation’, ZEA-3 unit at the Forschungszentrum Jülich, and the Institute for Inorganic and Analytical Chemistry at the TU Braunschweig, for the characterization of the polyethylene radiators.

\bibliographystyle{JHEP}
\bibliography{bibliography}

\end{document}